\documentclass[10pt]{article}

\usepackage{amssymb}
\usepackage{hyperref}
\usepackage{mathrsfs}
\usepackage{graphicx}% input figure files
\usepackage{dcolumn}% Align table columns on decimal point
\usepackage{bm}% bold math
\usepackage{mathrsfs}
\usepackage[utf8]{inputenc}
\usepackage{multirow}
\usepackage{booktabs}
\usepackage[perpage]{footmisc}
\usepackage{soul}
\usepackage[english]{babel}

\usepackage{color}
\usepackage{subfigure}
\usepackage{footnote}

\usepackage{chngcntr}
\usepackage[section]{placeins}
\usepackage{cite}

\begin{document}
\counterwithout{figure}{subsection}
\counterwithout{table}{subsection}

\title{Phase transitions in neutron stars and their links to gravitational waves}

\author{Milva G. Orsaria $^{1, 2}$, Germán Malfatti $^{1, 2}$, Mauro Mariani $^{1, 2}$,\\
Ignacio F. Ranea-Sandoval $^{1, 2}$, Federico García $^{4, 5}$, William M. Spinella $^{6}$,\\
Gustavo A. Contrera $^{1, 3, 8}$, Germán Lugones $^{7}$, and Fridolin Weber $^{8, 9}$}
\date{}
\maketitle
$^{1}$Grupo de Gravitaci\'on, Astrof\'isica y Cosmolog\'ia, Facultad de Ciencias Astron{\'o}micas y Geof{\'i}sicas, Universidad Nacional de La Plata, Paseo del Bosque S/N, 1900, La Plata, Argentina.\\
  $^{2}$ CONICET, Godoy Cruz 2290, 1425, CABA, Argentina.\\
  $^{3}$ IFLP, UNLP, CONICET, Facultad de Ciencias Exactas, Diagonal 113 e/63 y 64, 1900, La Plata, Argentina.\\
  $^{4}$ AIM, CEA, CNRS, Universit\'e Paris-Saclay, Universit\'e Paris Diderot, Sorbonne Paris Cit\'e, F-91191 Gif-sur-Yvette, France.\\
  $^{5}$ Instituto Argentino de Radioastronomía (CCT-La Plata, CONICET; CICPBA), C.C. No. 5, 1894 Villa Elisa, Argentina.\\
  $^{6}$ Department of Sciences, Wentworth Institute of Technology, 550 Huntington Avenue, Boston, MA 02115, USA.\\
  $^{7}$ Universidade Federal do ABC, Centro de Ciências Naturais e Humanas, Avenida dos Estados 5001- Bangú, CEP 09210-580, Santo André, SP, Brazil.\\
  $^{8}$ Department of Physics, San Diego State University, San Diego, CA 92182, USA.\\
$^{9}$ University of California at San Diego, La Jolla, CA 92093, USA.
%\ead{morsaria@fcaglp.unlp.edu.ar}

\begin{abstract}
The recent direct observation of gravitational wave event $GW170817$ and its $GRB170817A$ signal has opened up a new window to
study neutron stars and heralds a new era of Astronomy referred to as the Multimessenger Astronomy. Both gravitational and electromagnetic waves from a single astrophysical source have been detected for the first time. This combined detection offers an unprecedented opportunity to place constraints on the neutron star matter equation of state. The existence of a possible hadron-quark phase transition in the central
regions of neutron stars is associated with the appearance of g-modes, which are extremely
important as they could signal the presence of a pure quark matter core in the centers of neutron stars. Observations of g-modes with frequencies between 1 kHz and 1.5 kHz
could be interpreted as evidence of a sharp hadron-quark phase transition in the cores of neutron stars.
In this article, we shall review the description of the dense matter composing neutron stars, the determination of the equation of state of such matter, and the constraints imposed by astrophysical observations of these fascinating compact objects.
\end{abstract}


\section{Introduction}
\label{intro}
One of the principal aims of Hadron Physics is to provide a
comprehensive theoretical description of hadrons at the quark level,
which constitute the fundamental building blocks of matter.  The
hadrons are composite objects made up of three quarks (baryons) or of
quark-antiquark pairs (mesons). Light baryons are formed by different
combination of three light quarks, $u$ (up), $d$ (down), and $s$ (strange)\footnote{Despite the fact that heavy baryons and mesons composed
of $c$ (charm) and $b$ (bottom) quarks have been produced \cite{Tanabashi:2018oca},
we will ignore them. Quark chemical potentials inside neutron stars
are at most on the order of 500 MeV, too small to create a population of
these states.}. The gauge field theory with the local symmetry group SU(3) which describes the
strong interactions of colored quarks and gluons is known as Quantum
Chromodynamics (QCD).  In the framework of QCD, neutrons and protons,
which are the building blocks of ordinary atomic nuclei and of nuclear
matter, are the fundamental degrees of freedom of strongly interacting
matter at low temperatures and densities. At high temperatures and/or
densities, QCD predicts a phase transition of ordinary baryonic matter,
where quarks are confined inside of hadrons, to a new state of matter
called quark-gluon plasma (QGP), where the hadronic boundaries have disappeared and
quarks and gluons roam about freely.

It is understood that the universe was filled with such a plasma
immediately after the Big Bang. Moreover, it is being speculated that
quark matter may play a role in the demise of massive stars in
supernova explosions, could exist temporarily in the cores of (hot)
proto-neutron stars, and may form a permanent component of matter in
the cores of (cold) neutron stars (hereafter, NSs). It is therefore
extremely important to explore possible phase transitions in
super-dense matter, understand the thermodynamic behavior of the QGP, and to delineate realistic models for the
equation of state (EoS) of such matter.

Due to the property of asymptotic freedom the interactions between
quarks become increasingly weaker with decreasing separation.
This is the regime of high-momentum transfer (high densities and/or high temperatures),
where perturbative techniques may be used to determine the properties
of quark matter.  However, in the regime of low momentum transfer (low
densities and low temperatures, increasing separation), the strong coupling
constant becomes large and the theory is highly non-perturbative. In
this way, it is theoretically well-established that confinement and
chiral symmetry breaking cannot be obtained in a simple
perturbation-theoretical analysis of QCD. Thus, to describe hadron
masses or hadron decay constants it is essential to develop
formalisms that allows one to study the theory in the non-perturbative
region. A standard method to solve non-perturbative QCD analytically
does not exist. As an alternative, lattice QCD (LQCD)
\cite{Gattringer:2010zz} has been developed, which explores the
dynamics of quarks and gluons numerically on a discretized space-time
grid.  A very important consequence of the discretization is that the
measurement of the functional integrals are products of a large but
countable number of differentials.

Due to ever increasing computing power it has become possible to
decrease the spacing of the space-time grid and increase the size of
the system so as to approach both the continuum limit as well as the
thermodynamic limit. At the same time it has become possible to sample
a sufficiently large number of configurations in such numerical
simulations so that the statistical errors decrease
significantly. However, the Monte Carlo-type methods commonly used to
evaluate the partition function are reliable only in the case where
the chemical potential is zero. The extension of the calculations to
finite chemical potentials present serious difficulties (see Refs. \cite{Bellwied:2015rza,Ratti:2018ksb}, and references therein). A
path to treating QCD at finite chemical potential is to resort to
effective models of QCD, that is, models based on Lagrangians that are
simplified compared to the full QCD Lagrangian while maintaining the
fundamental properties of the full theory.

Studies of the QCD phase diagram predict that the crossover shown by
LQCD calculations at vanishing chemical potential becomes a
first-order phase transition at intermediate temperatures and high
baryon chemical potentials \cite{Schmidt:2017bjt}. In particular, this
suggests that the matter in the ultra-dense cores of NSs, which is
compressed to densities several times that of nuclear matter
($n_0 \sim$ 0.16 fm$^{-3}$), may undergo such a transition.

NSs are dense, neutron-packed remnants of massive stars that blew
apart in supernova explosions \cite{Weber:1999qn}. They are typically about 20 kilometers
across and spin rapidly, often making many hundred rotations per
second. The record holder among the rapidly spinning neutron stars
(pulsars) is currently PSR J1748-2446ad which rotates at 716 Hz
\cite{Hessels:2006}.

Many NSs are radio pulsars, emitting radio waves that appear
from the Earth to pulse on and off like a lighthouse as the star
rotates at very high speeds. NSs in X-ray binaries accrete material
from a companion star and flare to life with tremendous bursts of
X-rays. Pulsars can also be powered by extreme magnetic fields. Some
NSs, known as magnetars, have been shown to have surface magnetic
fields of 10$^{14}$-10$^{15}$ Gauss
\cite{Chakrabarty:1995br,Broderick_2000,Cardall_2001,Harding:2006qn}. It is believed that the magnetic fields in
their cores are even stronger by several orders of magnitude. From model
calculations it is known that the cores of NSs contain
most of their gravitational masses.  The nuclear EoS
associated with these regimes, however, is still largely unknown since
the description of the matter in the ultra-dense cores becomes
increasingly uncertain with density. This concerns both the many-body
techniques that are used to model such matter as well as
open issues regarding the compositional structure of the core, where
aside from neutrons and protons, hyperons, the delta resonance, meson
condensates, and/or deconfined quarks may exist
\cite{Weber:2000xd}. These features make NSs unique astrophysical
laboratories for a wide range of physical studies, which range from
the exploration of nuclear processes on the surfaces and inner crusts,
to the exploration of novel states of matter in the cores of such
objects.

Determining the NS EoS, that is, the functional relationship
between pressure and energy density, allows one to obtain the
relationship between the mass and radius of a NS as well as other key
properties such as the tidal deformability and the moment of
inertia. Strong constraints on the appearance of additional degrees of
freedom in neutron star matter, and hence on the EoS used to describe
such matter, follow from the discovery of NSs with masses of around
2M$_{\odot}$ \cite{Demorest:2010bx,Lynch:2012vv,Antoniadis:2013pzd,Arzoumanian_2018}. Moreover, important additional constraints on the
EoS of NSs follow from the recent direct detection of gravitational
waves (GWs) emitted from a merger of two NSs--the gravitational-wave
event GW170817--whose electromagnetic counterpart (GRB 170817A) has
been observed as well
\cite{TheLIGOScientific:2017qsa,GBM:2017lvd,Haggard:2017qne,Valenti_2017,Soares-Santos:2017lru,Abbott:2018wiz,Branchesi:2018ddy}. This event serves, for
instance, to explore the tidal deformability and radii of neutron
stars and to impose further constraints on the nuclear EoS
\cite{Annala:2017llu,Villar:2017wcc,De:2018,most2018}. Last but not least we mention
non-radial oscillation modes of NSs. Such modes may also lead to the
emission of GWs (see Refs. \cite{Benhar:2004xg,Andersson:2009yt,Ranea-Sandoval:2018bgu} and references therein), which could shed additional light on the behavior of the
matter in the interiors of NSs.

In this article, we shall review the composition of cold catalyzed
neutron star matter, the determination of the equation of state of
such matter, and the constraints imposed by astrophysical observations
of NSs. We shall focus on the nature and properties of matter at
supra-nuclear densities and low temperatures $T \leq 1$ MeV ($\simeq
10^{10}$ K). The article is organized as follows. In section \ref{SI}
we discuss the QCD phase diagram, effective models (and their
extensions) for the description of hadronic and quark matter, and the
constraints on quark matter imposed by LQCD studies. In section \ref{SII} we discuss the nature of the hadron-quarks phase transition. 
In
section \ref{SIII} we discuss constraints on the nuclear EoS from
radioastronomical observations of radio pulsars and X-ray observations of isolated NSs and X-ray binaries. In section \ \ref{SIV} we briefly describe the post GW170817 stage and discuss the NSs merger numerical simulation and the constraints to the EoS from the gravitational-wave event GW170817.  In section \ref{SV} we present some basic aspects of NSs oscillations and their link to GW emission from isolated NSs and theoretical results obtained from a representative hybrid
EoS. We discuss the relationship between a possible phase transition
in the inner core of a NS and GWs. We also provide an estimate of the
intensity of a GW signal linked to non-radial oscillation modes. A
summary and outlook is given in section \ref{SVI}. 

%Finally, in the
%Appendix  we provide the tabulated version of the nuclear
%EoS used in this work for the construction of hybrid stars (HSs), NSs with a phase of pure quark matter
%or a mixed phase with quarks in their inner core.

\section{Probing the dense matter EoS I. Terrestrial laboratories constraints}
\label{SI}

\subsection{Probing dense matter beyond nuclei} \label{QCD_generalities}

\begin{figure*}[htb]
\centering
\includegraphics[width=0.7\textwidth]{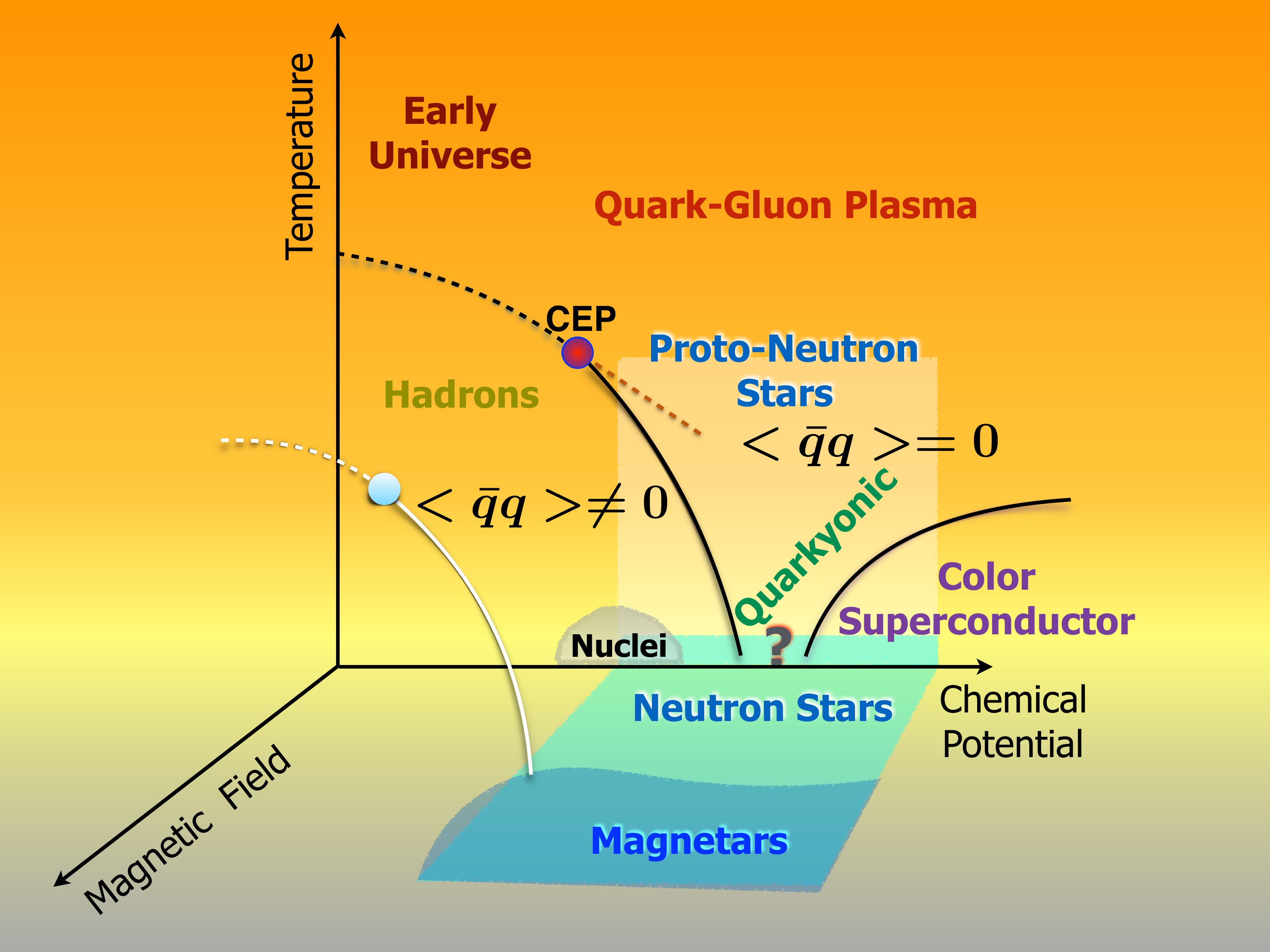}\\ \vspace{.5cm}
\caption{(Color online) Schematic drawing of the conjectured phase diagram of QCD as described in the text. The shaded region at low temperatures and high densities indicates where neutron stars and magnetars could lie according theoretical predictions.}
\label{diag_conj}
\end{figure*}

In the present review the focus is on the zone of the QCD phase diagram of high chemical potential and low temperatures. To study the behavior of matter subjected to such extreme conditions, as it could be found inside NSs, we should take into account the degrees of freedom relevant to low and high densities. From the theoretical point of view, it was originally believed that the QCD phase diagram consisted of only two phases: i) the hadronic phase of confined quarks and gluons with broken chiral symmetry, and ii) the deconfined phase with restored chiral symmetry, known as QGP. Only at the end of the 90's the possibility of new intermediate phases in the QCD phase diagram, like color superconductivity in which quark Cooper pairs are formed, began to be accepted \cite{Alford:2001dt} and later studied in detail in compact stars environments \cite{Alford:2007xm}. It has also been suggested that there may be non-homogeneous phases \cite{Carignano:2010ac,Buballa:2014tba,Carlomagno:2014hoa,Carlomagno:2015nsa}, and that there may be a phase of confined quarks with restored chiral symmetry (quarkyonic phase) \cite{McLerran:2007qj,McLerran:2008ua,Fukushima:2013rx}. A schematic version of the conjectured phase diagram of QCD is presented in figure \ref{diag_conj}.
According to LQCD simulations, a discontinuous chiral transition is expected at some critical end point (CEP) in the phase diagram of QCD, immediately after the crossover line in figure \ref{diag_conj}. The location and even the existence of the CEP are still open problems \cite{Ferreira2018}. 
Recent advances have been made in understanding the phase diagram of QCD in the presence of strong magnetic fields. The intensity of the magnetic field can affect the CEP location. Different model calculations and LQCD simulations predict magnetic catalysis at zero temperature (the quark condensate increases as a function of the magnetic field) \cite{RevModPhys88025001}. The so-called inverse magnetic catalysis in 2$+$1 flavors, in which the magnetic field suppresses the light $u$ and $d$ quark condensates close to the chiral restoration temperature, has been obtained using LQCD data \cite{PhysRevD.89.116011}. The phenomenon of inverse magnetic catalysis found in LQCD calculations was also verified theoretically considering two flavors of quarks \cite{PhysRevD.96.114012}. The nature of strong interactions under extreme external magnetic fields and its effect on the QCD phase diagram remains an open issue.

As mentioned in the introduction, the calculations of the LQCD are limited to
zero chemical potential for realistic quark masses as a consequence of the limitations of Monte Carlo techniques to evaluate the partition function. Due to the sign problem\footnote{The interpretation
of the canonical density operator as an evolution operator in imaginary time is
useful to study QCD at finite density, making the effective action a complex
quantity. This fact prevents comparing the probabilities associated with
different configurations.}, it is only possible to investigate the QCD phase
diagram at low chemical potentials developing theoretical alternative methods and using accurately extrapolation of the QCD data. Using a Taylor
expansion \cite{Allton:2002zi,Gavai:2008zr,Kaczmarek:2011zz} or analytical
continuation from imaginary chemical potential
\cite{deForcrand:2002hgr,DElia:2002tig,DElia:2007bkz,deForcrand:2008vr},
the behavior of the transition temperature at finite density in
the low density region of the QCD phase diagram has been studied (see
Ref. \cite{Bellwied:2015rza} for a complete review of this issue). At zero chemical
potential some observables (chiral condensate and its susceptibility) that are
related to the spontaneous chiral symmetry breaking give a critical temperature
around $T_c(\mu = 0)\simeq 155$ MeV, obtained from extrapolations to the continuum limit
for $2+1$ quark flavors \cite{Bazavov:2011nk,Aoki:2009sc,Bazavov:2016uvm}. In
addition, some experimental data from the Beam Energy Scan at RHIC by the STAR
collaboration and by the ALICE Collaboration at the LHC show some signals of
deconfinement phase transition from the determination of the chemical
freeze-out temperatures and baryochemical potentials
\cite{Cleymans:2004pp,Becattini:2005xt,Andronic:2005yp,Andronic:2008gu,Alba:2014eba}.

Although the low density regime of the QCD phase diagram can be
probed by the LQCD calculations through the approaches mentioned before, the formulation of LQCD at higher values of the chemical
potential fails. Thus, phenomenological approaches are necessary to interpret the recent experimental results at NICA-FAIR energies \cite{Bugaev:2018lfj}. These approaches basically consist on the assumption the production of new degrees of freedom in the early stage of nucleus-nucleus collisions is a statistical process. The broad set of experimental data obtained at NICA-FAIR energies agrees with the interpretation of the results using statistical models based on such statistical process. This enables to extract information from the transition zone between the hadronic and quark-gluon phases at the region of the QCD phase diagram covered by these experiments.

Due the separation of the energy scales in the treatment of strong interactions (low and high energy regimes), powerful tools to study the dense matter behavior are effective theories that keep some of the features of the QCD. The most popular effective models used for hadron
interactions are the Nambu--Jona Lasinio (NJL) models
\cite{Vogl:1991qt,Klevansky:1992qe,Hatsuda:1994pi,Buballa:2003qv}.
Other models, like the Chiral Sigma Model
\cite{GellMann:1960np,Ketov:2000dy,Mizher:2010zb,Meurice:2017zng}, the
Polyakov--Quark-Meson Model
\cite{Schaefer:2004en,Schaefer:2007pw,Zacchi:2015oma,Sedrakian:2017qpg}, or the
Field Correlator Method
\cite{Dosch:1998th,Simonov:2005jj,Nefediev:2009kn,Plumari:2013ira,Mariani:2016pcx}
have also been used to describe the quark matter phase.

So far, theoretical physicists have not managed to find an effective QCD model
capable of reproducing both the hadron and quark regimes in a single EoS. Matter
becomes simpler to describe at very high densities due to the asymptotic
freedom property, but very complicated in the opposite limit due to the
confinement. What happens in the middle is speculation that comes from
combining the data extracted from heavy ion collisions and experimental physics
with different theoretical models.

NSs can provide a unique window to explore strong interactions under extreme
conditions. Thus, to model the matter in the inner and outer cores of NSs,
where the bulk of their mass is concentrated, theoretical astrophysicists
consider two distinct phases composed of different types of particles. In this
way, a hybrid EoS will be a multicomponent system composed of two independent
phases: the hadronic phase at low densities, and the quark phase at high
densities.

\subsection{Hadronic phase at low densities} \label{Hadron_phase}

To describe the interaction between baryons, we use Quantum Hadrodynamics, the relativistic quantum field theory of interacting many-body systems based on hadronic degrees of freedom. Such description can be achieved by modeling the interaction between two baryons as a coupling of the baryons to meson fields, which reproduce the fundamental attractive and repulsive nature of the strong force. This approach is similar to the one-boson-exchange modeling of the nuclear interaction that very successfully describes the wealth of nucleon-nucleon scattering data. However, instead of fixing the meson-nucleon coupling constants to scattering data, they are fit to the properties of bulk nuclear matter at the nuclear saturation density extracted from recent experimental or theoretical results from nuclear physics.

A popular model of Quantum Hadrodynamics 
is known as the Walecka model \cite{Walecka:1974qa,Boguta-Bodmer1977,Serot:1984ey} in which all the baryon states, $B$,  (nucleons, hyperons which possess strange quarks and delta resonances)
interact via the exchange of scalar, vector and isovector mesons ($\sigma $,
$\omega$, $\rho$, respectively). In this model, the meson field strengths are
taken to be equal to their mean values in the so called relativistic mean-field (RMF) approximation, and 
the baryon-baryon interactions are described in terms of meson fields. Thus, the scalar meson $\sigma$ describes attraction between baryons, the vector meson $\omega$ describes
repulsion, and the isovector meson $\rho$ is necessary for describing baryon-baryon interactions in isospin asymmetric systems. The EoS of isospin asymmetric nuclear matter is crucial in
understanding the structure of NSs, specially for their cooling history \cite{Lattimer:1994glx}.

The Lagrangian of nuclear matter constructed
on the basis of the
original Walecka model and extended by the inclusion of nonlinear scalar
self-interaction terms is given by \cite{Glendenning:1997wn},
\begin{eqnarray}
  \mathcal{L}_{B} &=& \sum_{B}\bar{\psi}_B \bigl[\gamma_\mu [i\partial^\mu - g_{\omega B}(n)
    \omega^\mu - g_{\rho B}(n) {\boldsymbol{\tau}} \cdot {\boldsymbol{\rho}}^\mu] - [m_B - g_{\sigma B}(n)\sigma]
    \bigr] \psi_B  \nonumber \\
&&+ \frac{1}{2} (\partial_\mu \sigma\partial^\mu
  \sigma - m_\sigma^2 \sigma^2) - \frac{1}{3} b_\sigma m_N [g_{\sigma N}(n)
  \sigma]^3 - \frac{1}{4} c_\sigma [g_{\sigma N}(n) \sigma]^4 \nonumber \\
&&- \frac{1}{4}\omega_{\mu\nu} \omega^{\mu\nu}
  +\frac{1}{2}m_\omega^2\omega_\mu \omega^\mu + \frac{1}{2}m_\rho^2
  {\boldsymbol{\rho\,}}_\mu \cdot {\boldsymbol{\rho\,}}^\mu - \frac{1}{4}
  {\boldsymbol{\rho\,}}_{\mu\nu} \cdot {\boldsymbol{\rho\,}}^{\mu\nu} \, , \label{eq:Blag}
\end{eqnarray}
where $g_{\sigma B}(n)$, $g_{\omega B}(n)$ and $g_{\rho B}(n)$ are the (density
dependent) meson-baryon coupling constants, $n$ is the baryon number density,
${\boldsymbol{\tau}}$ are the Pauli isospin matrices, and $\gamma^{\mu}$ are the Dirac
matrices.

\begin{table}[tb]
\begin{center}
  \caption{Properties of nuclear matter at saturation
  density for the hadronic DDRMF parametrizations, GM1L \cite{Spinella2017:thesis} and DD2 \cite{Typel:2009sy}.}
\begin{tabular}{ccc}
\hline\noalign{\smallskip}
$~~$Saturation Property$~~$ & $~~$GM1L$~~$
&DD2$~~$\\
\noalign{\smallskip}\hline\noalign{\smallskip}
$n_0$  (fm$^{-3}$)    & 0.153      & 0.149     \\
$E_0$  (MeV)          & $-16.30$    & $-16.02$  \\
$K_0$  (MeV)          & 300.0        & 242.7    \\
$m^*/m_N$             & 0.70       & 0.56    \\
$J$    (MeV)          & 32.5         & 32.8     \\
$L_0$  (MeV)          & 55.0         & 55.3    \\
\noalign{\smallskip}\hline\noalign{\smallskip}
\end{tabular}
\label{table:parametrizations}
\end{center}
\end{table}
In the standard RMF the meson-nucleon
coupling constants, $g_{i N}(n) = g_{i N}(n_0)$,  $ i \in \{\sigma,
\omega, \rho\}$ and the coefficients of the nonlinear scalar self interactions
$b_\sigma$ and $c_\sigma$, are set so as to reproduce the properties of isospin
symmetric nuclear matter. Symmetric nuclear matter, $n_{p} = n_{n}$, is a
reasonable approximation for the interior of heavy atomic nuclei and 
can be parameterized to reproduce empirically determined
properties of nuclear matter at the nuclear saturation density, $n_{0}$,
thereby constraining the EoS.  These saturation properties are the binding
energy per nucleon ($E_0$), the nuclear incompressibility ($K_0$), the isospin
asymmetry energy ($J$) and its slope ($L_0$), and the effective mass
($m^{*}/m_N$), all fixed to their $n_{0}$ values. However, the standard RMF
approximation is not parameterized to fix $L_0$ at the saturation density, a
quantity that has become more tightly constrained in recent years and may have
a significant effect on the composition and properties of NSs (see Refs. \cite{Lattimer:2014sga,Lopes:2017ovz} and references therein). One of the ways to set $L_0$ is
to consider medium effects of the effective interaction, choosing density
functionals for the meson-baryon couplings.  More specifically, the density
dependence is extracted from Dirac-Brueckner calculations in which nucleon
interactions in the nuclear medium are obtained from nucleon-nucleon potentials
consistent with scattering experiments \cite{RocaMaza:2011qe}. Being based on
first principles calculations, density dependent coupling constants give more
empirical robustness to the density-dependent relativistic mean-field (DDRMF)
than in the standard RMF.  
The functional forms for the density dependent
isovector- , scalar- and vector-meson-baryon coupling constants
are given by \cite{Typel:1999yq,Typel:2018cap}
\begin{eqnarray*}
g_{\rho B}(n) = g_{\rho B}(n_0)\,\, \mathrm{exp}[\,-a_{\rho} (x - 1)\,],\\
\vspace{2.0cm}
g_{i B}(n) = g_{i B}(n_0)\,\, a_{i}\,\, \frac{1+b_{i}(x+d_{i})^{2}}{1+c_{i}(x+d_{i})^{2}},\quad\mathrm{for}\quad
i=\sigma,\omega,
\end{eqnarray*}
where $x = n/n_0$. This choice of parametrization accounts for
nuclear medium effects by making the meson-baryon coupling constants dependent
on the local baryon number density \cite{Fuchs:1995as}. The parameters $a_{i}$,
$b_{i}$, $c_{i}$ and $d_{i}$ are adjusted by comparing model predictions
to experimental data such as binding energies, charge and diffraction radii,
spin-orbit splittings and neutron skin thickness of finite nuclei
\cite{Typel:2005ba}. The density dependence of the meson-baryon couplings
also eliminates the need for the non-linear self-interactions in the DDRMF
Lagrangian. Additionally, as a consequence of the density dependent coupling, a
rearrangement contribution term in the chemical potential and pressure
is mandatory to maintain thermodynamical consistency \cite{Hofmann:2000mc}.

The determination of the meson-hyperon coupling constants at the saturation density is more complicated. Since hyperons are not present in the atomic nuclei, it is not possible to adapt the meson-hyperon coupling constants to any property of the nuclear matter as in the case of nucleons. To scape this problem, the simplest proposal is to suppose that the coupling constants have the same value for all baryons. This is the so-called universal coupling. Another alternative is to determine the meson-hyperon coupling constants based on the spin-isospin flavor symmetry SU(3) \cite{Spinella2017:thesis}. 

The
saturation properties associated with the DDRMF parameterizations are given in
table \ref{table:parametrizations}. Results using EoS calculated employing the
GM1L and DD2 parametrizations are shown and discussed in sections \ref{SIV} and \ref{SV}.

\subsection{Quark phase at low, intermediate and high densities} \label{Quark_phase}

As it has been mentioned in \ref{QCD_generalities}, to study the physical
processes among quarks in the low momentum range of QCD, it is possible to use
effective models that are described by a Lagrangian that accounts for the
main features of QCD at low energies. For example, NJL-like models describe
interactions between constituent quarks and give a simple scheme to study
spontaneous chiral symmetry breaking, a key feature of QCD in its low
density/temperature phase, and its manifestations in hadronic physics, such as
dynamical quark mass generation, the appearance of a quark pair condensate and
the role of pions as Goldstone bosons. In its simplest form, the NJL model uses
only a local scalar-isoscalar and pseudoscalar-isovectorial interaction between
fermions. The problem with such local interactions is, however, that the model
should be regularized in order to avoid divergences in quantities such as the
quark self-energies and the meson masses. In the 90's, it was suggested to
use non-local interactions rather than local interactions \cite{Ripka:332104}. A
natural way to introduce non-locality in the quark-quark interaction is to use
a phenomenologically modified (effective) gluon propagator. These modified propagator leads to a non-local regulator for quark currents that have the advantage to decay at large momentum transfer, thus enabling to numerically perform the integrals to infinity without using a momentum cut-off. In general,
these models lack confinement, which is one of the key properties
of QCD. The Polyakov loop dynamics is a suitable framework to achieve quark
confinement in these kind of models, in particular at finite temperatures and
vanishing densities.

Polyakov-NJL (PNJL) models have an effective potential ${\cal{U}}(\Phi ,T)$
defined for the (complex) $\Phi$ field, which is conveniently chosen to
reproduce, at the mean field level, results obtained in LQCD calculations.
Different ansatz for $\cal{U}$ are available in the literature (see, for example, Refs. 
\cite{Roessner:2006xn,Ratti:2005jh,Fukushima:2008wg}). The effective potential
exhibits the feature of a phase transition from color confinement ($T < T_0$,
the minimum of the effective potential being at $\Phi = 0$) to color
deconfinement ($T > T_0$, the minimum of the effective potential occurring at
$\Phi \neq 0$), where the parameter $T_0$ is the critical temperature for the
deconfinement phase transition. This parameter is the only free parameter of
the Polyakov loop once the effective potential is chosen. If a nonlocal
extension of the SU(3) PNJL model is considered, the Lagrangian can be written
as
\cite{Contrera:2007wu,Contrera:2009hk,Contrera:2010kz,Contrera:2012wj,Carlomagno:2013ona,Malfatti:2017cln,Carlomagno:2018tyk}:
\begin{eqnarray}
\mathcal{L} &=&  \bar \psi (x) \left[ -i\gamma_{\mu}\partial_{\mu} + \hat m \right] \psi(x) - \frac{G_s}{2}
\left[ j_a^S(x) \ j_a^S(x) + j_a^P(x) \ j_a^P(x) \right]
\nonumber \\&& - \frac{H}{4} \ T_{abc} \left[
  j_a^S(x) j_b^S(x) j_c^S(x) - 3\ j_a^S(x) j_b^P(x) j_c^P(x)
  \right]+ \ {\cal U}
\,(\Phi,T)\;,
\label{eq:L_E}
\end{eqnarray}
where $j_a^{S,P}(x)$ are the scalar and pseudoscalar currents given by

\begin{eqnarray}
j_{a}^S(x) &=& \int d^4z \tilde{Rg}(z)\bar \psi \left(x + \frac{z}{2}
\right)\lambda_a \psi \left(x - \frac{z}{2} \right) \nonumber
\\ j_{a}^P(x) &=& \int d^4z \tilde{Rg}(z)\bar \psi \left(x +
\frac{z}{2} \right) i\lambda_a \gamma^5 \psi \left(x - \frac{z}{2}
\right) ,
\end{eqnarray}
with the regulator $\tilde{Rg}(z)$ written into momentum space, $Rg(p) =
\mathrm{exp}(-p^2/\Lambda^2)$, and $\Lambda$ a cutoff 
adjusted to set the range of the non-locality.
In this model, the quark current masses and coupling constants have be chosen
so as to reproduce the phenomenological values of pion decay constant $f_\pi$,
and the meson masses $m_{\pi}$, $m_{\rm K}$ and  $m_{\eta'}$. Setting the strange
quark mass to the updated value of $m_s=95$ MeV \cite{Tanabashi:2018oca}, the
parameters $\Lambda = 1071.38$ MeV, $G_s \Lambda^2 = 10.78$, $H \Lambda^5 =
-353.29$ and $m_u=m_d=3.63$ MeV \cite{Malfatti:2017cln} are obtained. 
\begin{figure*}[htb]
\centering
 \includegraphics[width=0.7\textwidth]{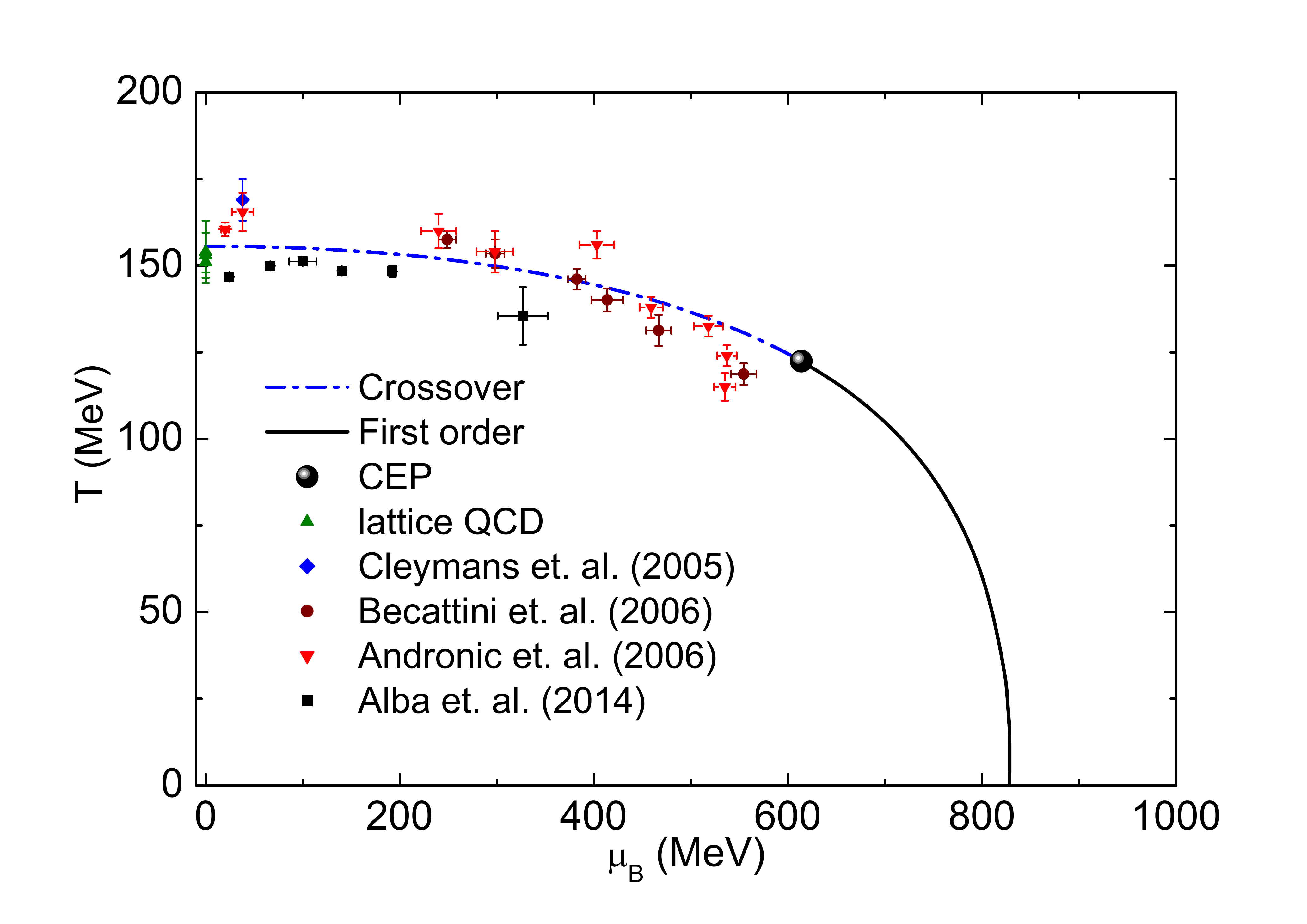}
\caption{(Color online) Phase diagram obtained with the non-local PNJL model,
  lattice QCD results \cite{Bazavov:2011nk,Aoki:2009sc,Bazavov:2016uvm} and
  chemical freeze-out from relativistic heavy ion collision data
  \cite{Cleymans:2004pp,Becattini:2005xt,Andronic:2005yp,Andronic:2008gu,Alba:2014eba}.
  The solid (dot-dashed) line corresponds to the first order (crossover) phase
  transition, and the solid dot represents the location of the obtained CEP.}
  \label{freezeout}
\end{figure*}
To extend the model at finite temperature and study different thermodynamical quantities, the
Matsubara imaginary time formalism is used \cite{Kapusta:2006pm}. Once the
grand canonical potential is obtained for the mean field approximation after the
bosonization of equation \ref{eq:L_E}, the phase diagram of figure
\ref{freezeout} can be calculated, where the traced (blue) line is a crossover
phase transition defined by the critical points having a peak in their chiral
or baryonic susceptibility, the solid (black) line indicates the first order
phase transition defined by the critical points where the two minims of the
grand canonical potential are equal, and the solid dot indicates the location
of the critical end point (CEP) where the first order phase transition line
ends and a second order phase transition occurs. Freeze-out and lattice data
points mentioned previously are shown in figure \ref{freezeout}.
These data points have also been compared to several analytical extensions of
LQCD in \cite{Bellwied:2015rza}. It is worthly to remark that in both local and nonlocal extensions of NJL models there are absence of a hadronic phase of confined quarks. The overlayed chemical freeze-out data in the phase diagram of the non-local model are presented only by completeness.  Chemical freeze-out parameters are analyzed from particle ratios in heavy-ion collision experiments using the statistical model of the hadron resonance gas \cite{Ratti:2018ksb}. The non-local PNJL model at zero temperature
has been used to describe quark matter in compact stars interiors in several
previous works
\cite{Orsaria:2012je,Orsaria:2013hna,Ranea-Sandoval:2017ort,Ranea-Sandoval:2015ldr}.

Recently, the Field Correlator Method, a non-perturbative approximation of QCD
parametrized through the gluonic condensate and the quark-antiquark static
potential for long distances (string confinement), has also been used as an alternative effective
model to study quark matter
\cite{Plumari:2013ira,Mariani:2016pcx,Simonov:2009nf,Burgio:2015zka,
Alford2015}.
\begin{figure}[!htbp]
\centering
\includegraphics[width=.8\textwidth]{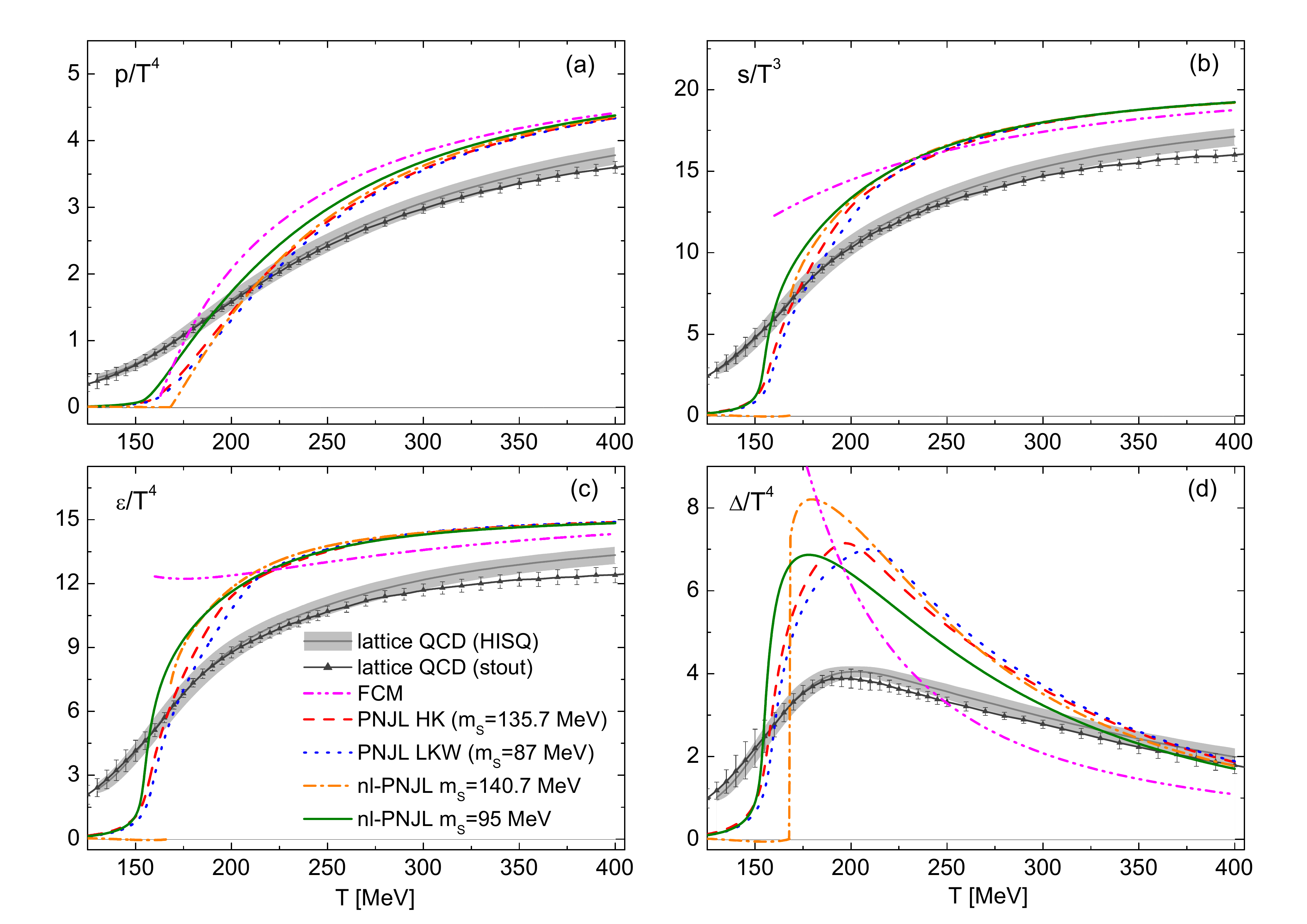}
\caption{(Color online) Comparison of some of thermodynamic properties of the effective quark models with continuum extrapolations of LQCD, as described in the text.}
\label{comp_lqcd}
\end{figure}

Some thermodynamical functions can be determined from the standard derivation
of the grand canonical potential. In figure \ref{comp_lqcd} the results
obtained with the non-local SU(3) PNJL model are compared to the continuum
extrapolation for 2+1 flavors in LQCD obtained with the stout action by the
Wuppertal-Budapest Collaboration \cite{Borsanyi:2013bia} and the results of
HISQ Action obtained by HotQCD Collaboration \cite{Bazavov:2014pvz}. For
completeness the calculations have been performed using several model
parameterizations which consider different ranges of strange quark masses, for
local (Hatsuda and Kunihiro (HK) \cite{Hatsuda:1994pi} and Lutz, Klimt, and
Weise (LKW) \cite{Lutz:1992dv}) and non-local (for $m_s=140.7$ MeV
\cite{Orsaria:2012je,Orsaria:2013hna} and $m_s=95$ MeV \cite{Malfatti:2017cln})
PNJL models. The results obtained from the FCM has also been included.

Although the existence of color superconductivity had been suggested in the
mid-70's \cite{Barrois:1977xd}, the possibility of this alternative phase of
quark matter at low temperatures and high densities was mostly accepted twenty
years later. Then, it was argued that its ground state
consists of quark pairs which form a color superconductor where the attractive
interaction between quarks emerges naturally through the exchange of gluons \cite{Alford:2001dt}. As
seen in the Bardeen-Cooper-Schrieffer (BCS) theory of ordinary
superconductivity, quarks with equal but opposite momenta at their respective
Fermi surfaces form Cooper pairs to lower the energy of the system. Color
superconductivity in the interiors of compact stars
has been studied by many authors over the last 15
years \cite{Sedrakian:2017qpg,Alford2003,BALDO2003,Bombaci:2006cs,Klahn:2006iw,Ippolito2008,Lugones:2009ms,Lugones:2010gj,Paulucci2011,Bonanno:2011ch,Ayvazyan:2013cva}.

In particular, color superconducting NJL 2SC+s quark matter cores in HSs has been studied in Ref. \cite{Ranea-Sandoval:2017ort}. 
In the NJL 2SC+s phase, green and red up and down quarks form pairs and are
embedded in a gas of free strange quarks. Results using this high
density EoS are shown and discussed in section \ref{SIV}.

\section{Dense hybrid-matter in NSs structure}
\label{SII}
\begin{figure*}[htb]
\centering
\hspace*{\fill}%
       \subfigure[Hypothetical NS cross section with a pure quark phase.]{%
            \label{fig:maxwell-diagram}%
          \includegraphics[width=0.44\textwidth]{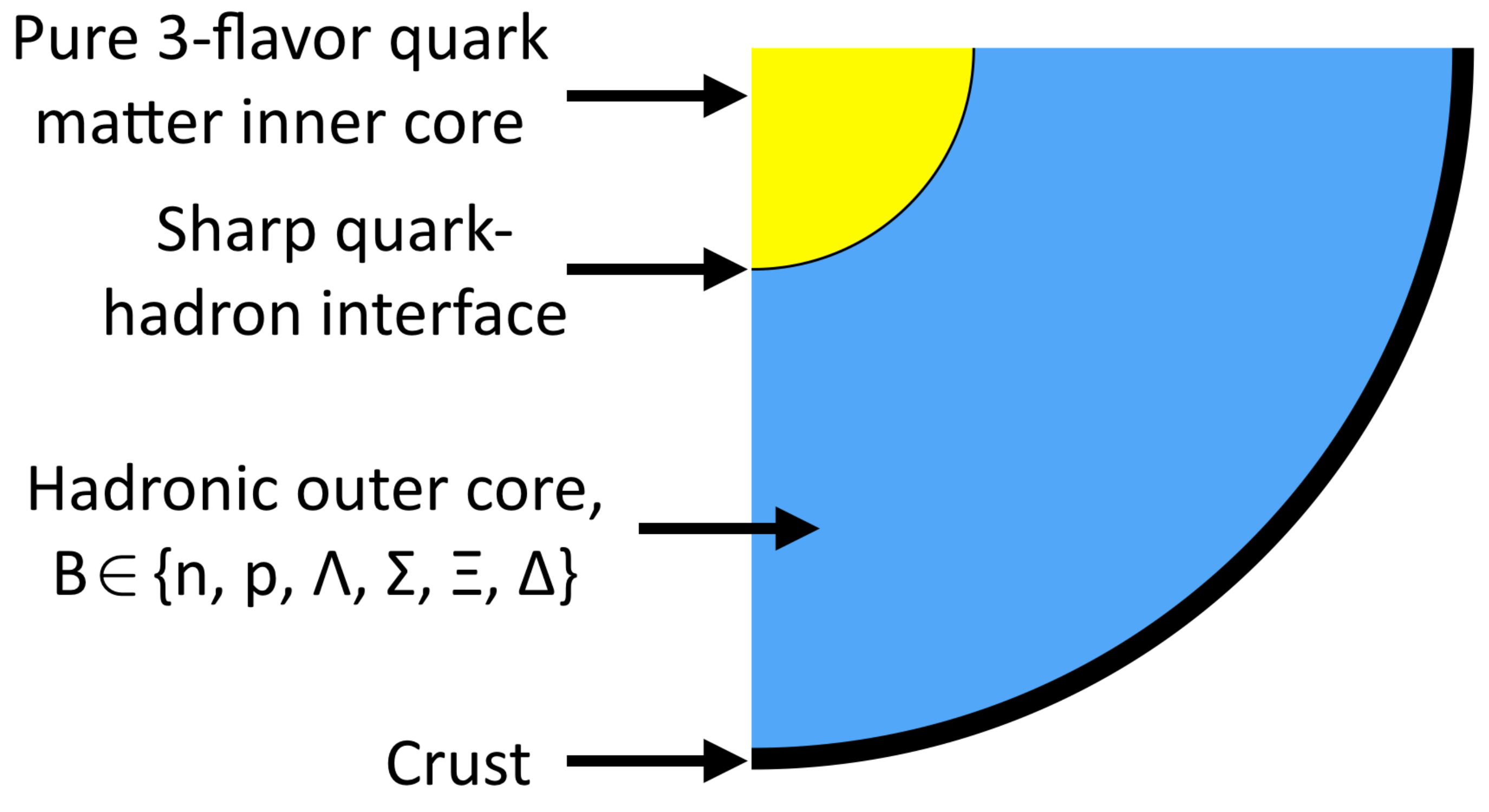}} \hfill%
       \subfigure[Hypothetical NS cross section with a mixed phase.]{%
           \label{fig:gibbs-diagram} %
        \includegraphics[width=0.41\textwidth]{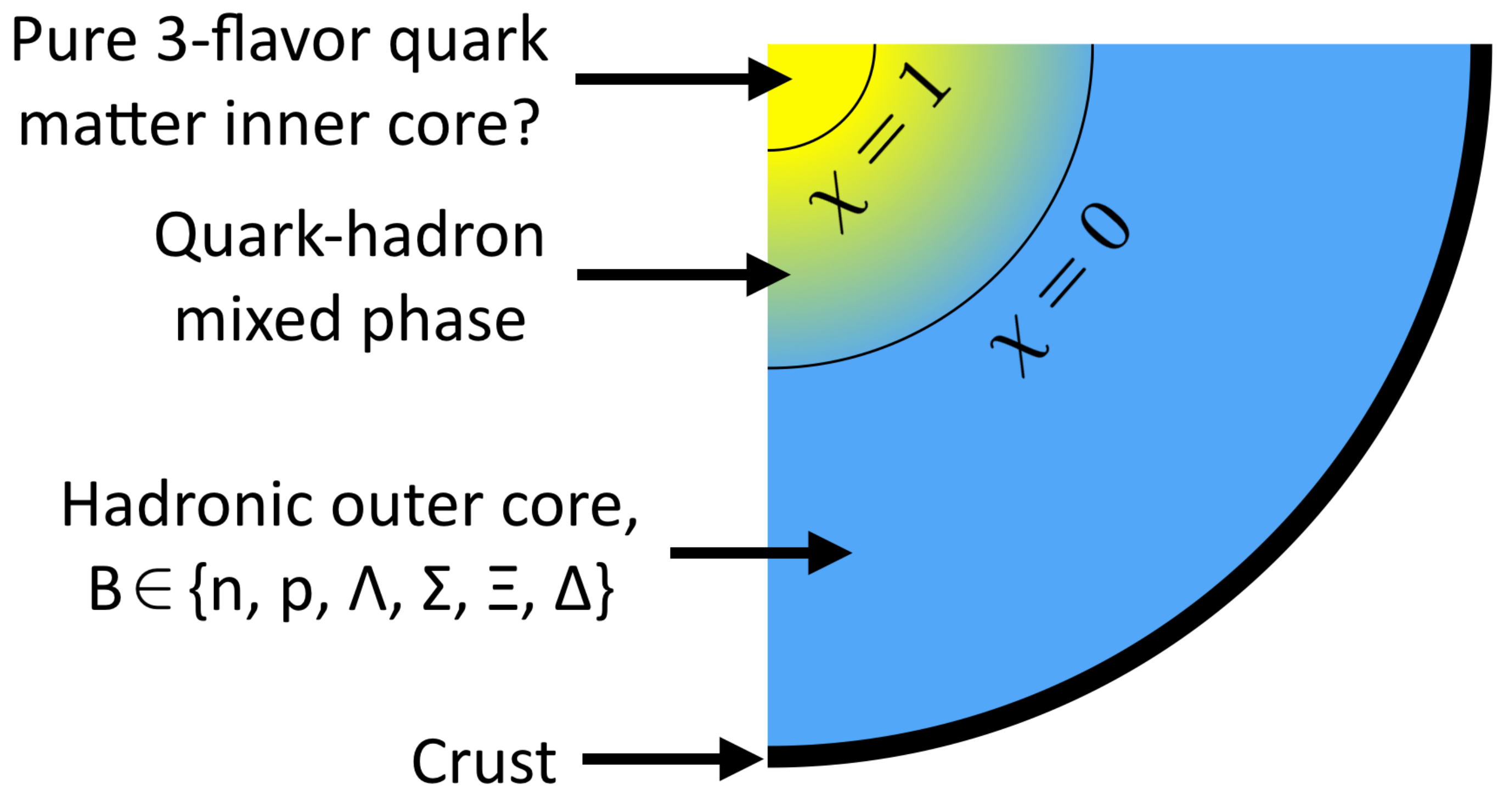}} %
\hspace*{\fill}%
\caption{(Color online) Hypothetical NS cross section assuming a constant
  pressure for a Maxwell \ref{fig:maxwell-diagram} or Gibbs \ref{fig:gibbs-diagram}
  construction phase transition from hadronic matter to a pure quark matter
  core \cite{Spinella2017:thesis}.}
  \label{diagrams2}
\end{figure*}

Different EoSs define different types of stars: NSs, HSs,
and strange stars \cite{Weber:2004kj}. Allowing for the possibility of quark deconfinement in NSs, a
hadron-quark phase transition may commence when the pressure of the quark phase
equals that of the hadronic phase. The nature of the phase transition depends
on the hadron-quark surface tension, $\sigma_{HQ}$, between the two phases that
is still quite uncertain.  Recent works have typically placed $\sigma_{HQ}
\lesssim 30$ MeV/fm$^2$, though there are suggestions that values greater
than 100 MeV/fm$^2$ could be possible too
\cite{Palhares:2010be,Pinto:2012aq,Mintz:2012mz,Lugones:2013ema}. If
$\sigma_{HQ} \gtrsim 70$ MeV/fm$^2$ the quark-hadron phase transition will be
one of constant pressure with an equation of state that is discontinuous in
energy density \cite{Voskresensky:2002hu,Yasutake:2014oxa}. The result is a
sharp interface between phases of pure hadronic matter and pure quark
matter at a given NS radius, as shown schematically in figure
\ref{fig:maxwell-diagram}.

The phase transition is achieved by applying the Maxwell construction
(at zero temperature, $T$) and imposing $P_H$ ($\mu_{Bt},T=0$) = $P_Q$ ($\mu_{Bt},T=0$),
where $\mu_{Bt}$ is the baryon chemical potential at the transition. $P_H$ and $P_Q$
represent the pressure of the hadronic and the quark phase respectively. The condition for a stable phase 
transition from hadron to quark matter is fulfilled if the Gibbs free energy of quark phase is less than the Gibbs free energy of the hadron phase, for pressures above the transition pressure.
If $\sigma_{HQ} \lesssim 70$ MeV/fm$^2$ the
phase transition results in the formation of a stable coexistent
(mixed) phase, gradually converting NS matter from hadronic matter to
deconfined quark matter with increasing density, as shown in figure
\ref{fig:gibbs-diagram}.
This phase transition satisfies the Gibbs condition for phase
equilibrium, $P_H$($\mu_n$,$\mu_e,T=0$) = $P_Q$($\mu_n$,$\mu_e,T=0$),
where $\mu_e$ is the electron chemical potential. The isospin
restoring force favors a positively charged hadronic phase as an
increased number of protons reduces isospin asymmetry.  This
necessarily results in a negatively charged quark phase, electric
charge neutrality being achieved globally.

\begin{figure}[htb!]
\begin{center}
\includegraphics[scale=0.5]{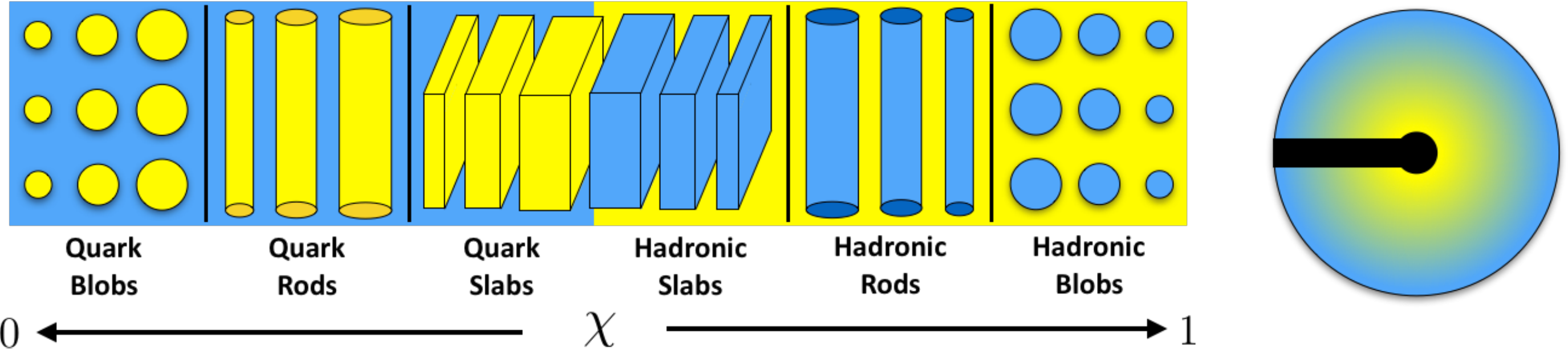}
\caption{Schematic illustrating the rare phase structures that may form in the
  mixed phase. An increase in $\chi$ is accompanied by an increase in baryonic
  number density within the NS. Figure adapted from Ref. \cite{Spinella2017:thesis}.}
   \label{fig:shapes}
\end{center}
\end{figure}

The presence of quark matter enables the hadronic regions of the mixed phase to
arrange themselves to be more isospin symmetric than in the purely hadronic
phase by transferring
charge to the quark phase in equilibrium with it. The symmetry energy will be
lowered thereby at only a small cost in rearranging the quark Fermi surfaces.
The electrons play only a minor role when neutrality can be achieved among the
baryon-charge carrying particles. The stellar implication of this charge
rearrangement is that the mixed phase region of the star will have positively
charged regions of nuclear matter and negatively charged regions of quark
matter. Because of the competition between the Coulomb and the surface energies
associated with the nuclear and quark regions, the mixed phase may develop geometrical
structures, schematically illustrated in figure \ref{fig:shapes}, 
as is similarly expected of the sub-nuclear liquid-gas phase
transition \cite{Ravenhall:1983uh,Williams:1985prf}. The quantity $\chi \equiv
V_q/V_{TOTAL}$ in such figure denotes the volume proportion of quark matter,
$V_q$, in the mixed phase. By definition, $\chi$ therefore varies between 0 and
1, depending on how much hadronic matter has been converted to quark matter.

Competition between the Coulomb and surface energies establishes the shapes,
sizes, and spacings of the rare phase in the background of the other in order
to minimize the lattice energy \cite{Na:2012td}. The change in energy
accompanied by developing such geometrical structures is likely to be very
small in comparison with the volume energy \cite{Glendenning:2001pe} and, thus,
may not affect the global properties of a NS. However, the
geometrical structure of the mixed phase may be very important for
irregularities like pulsar glitches in the timing structure of pulsar spin-down
as well as for the thermal and transport properties of NSs
\cite{Spinella:2015ksa}. The effect of a crystalline quark-hadron mixed phase
on the neutrino emissivity from the cores of NSs has been recently studied in Refs. \cite{Spinella:2015ksa,Spinella:2018bdq}.

Summarizing, the phase transition that might occur between the inner and outer
cores of a NS could be described within different scenarios, depending on the
value of $\sigma_{HQ}$. There are two limiting cases, the Maxwell construction,
which assumes $\sigma_{HQ} \to \infty$ and the {\it bulk} Gibbs construction,
which assumes $\sigma_{HQ} \to 0$. For intermediate values of $\sigma_{HQ}$,
the {\it full} Gibbs formalism, {\it i.e.} a full pasta phase calculation, must
be used taking into account the formation of phase structures in the mixed
phase \cite{Wu:2017xaz}. Recently, several works have included finite-size effects at the hadron-quark phase transition in a phenomenological way, mimicking the mixed phase by a function which depends on the broadening of the phase transition \cite{Ranea-Sandoval:2018bgu,Yasutake:2014oxa,2018Univ....4...94A}.  Therefore, the Gibbs construction for the mixed phase can be simulated through a continuous interpolation between the hadronic and quark phases to mimic the mixing or percolation \cite{Ranea-Sandoval:2018bgu}. Results using a sharp phase transition are shown in section \ref{SIV}.

\section{Probing the dense matter EoS II. Astronomical constraints pre GW17081}
\label{SIII}

NSs were first observed serendipitously in the radio wavelengths as {\em pulsars}: extremely stable periodic signals that spin-down slowly with high regularity as they loose rotational energy through dipolar radiation. Soon after their discovery, radio pulsars were interpreted as rapidly-rotating highly-magnetized NSs. Currently, $\sim$2000 pulsars are known\footnote{See http://www.atnf.csiro.au/research/pulsar/psrcat}. Despite their precise temporal behavior, more than 500 glitches in about 180 pulsars have been detected \cite{Manchester:2018jhy}. Those glitches are characterized by sudden changes in the spin period, followed by a gradual relaxation phase. They could be originated during the fast coupling of the spin rates of the crust and the core of the NS. Continuous pulsar monitoring at radio wavelengths can give clues about the NS interior composition through timing analysis of glitching activity \cite{Espinoza:2011pq,fuentes2017}. 

NSs are also observed in the X-ray band. Most of them are found in X-ray binaries (XRBs) where the X-ray radiation is produced when matter from the companion stars is accreted by the NSs. Isolated NSs have also been detected in X-rays, either young NSs in supernova remnants, such as the Crab and Vela pulsars, but also middle-aged close-by NSs, such as the famous Magnificent Seven \cite{Kaplan:2008qn}. Some NSs have also been detected in the UV-optical bands and more recently in the near infrared \cite{Posselt:2018ndf}.

The information gathered from pulsar timing of glitches and X-ray spectral analysis of NSs showing thermal cooling suggests that a hadronic crust must be present in at least the vast majority of the NSs observed. Those crusts are essentially composed by neutron-rich nuclei forming an ordered lattice, and a gas of free neutrons and electrons in beta equilibrium. NSs in X-ray binaries are able to accumulate new, light material on their hard surfaces that burns through unstable thermonuclear explosions known as X-ray bursts. Detailed temporal and spectral studies of the X-ray emission of NSs can lead to an estimation of their masses, radii, moments of inertia and internal composition. Several more involved reviews focused on the observational achievements and issues encountered can be found in Refs. \cite{Lattimer:2012nd,Ozel:2012wu,Miller:2013tca,Lattimer:2015nhk,Miller:2016pom}.
We now describe some of the most important astrophysical results previous to the discovery of GW170817.

Besides the two NSs involved in GW170817, there are about 40 NSs with estimated masses\footnote{See https://stellarcollapse.org/nsmasses}. Some of them are extremely precisely determined from their orbital parameters. NSs masses tend to concentrate around the canonical 1.4~M$_\odot$ but the most stringent observational constraints on the NS EoS arise from two pulsating NSs in binary systems with white dwarfs companions: PSR J1614--2230 \cite{Demorest:2010bx} with $M=1.908\pm0.016$~M$_\odot$ \cite{Arzoumanian_2018}, and PSR J0348+0432, with $M=2.01\pm0.04$~M$_\odot$ \cite{Antoniadis:2013pzd}. These highly precise measurements based on timing analysis, challenge the high-density EoS models, but also impose limits on the EoS at lower densities (see figure \ref{masa_radio} in section \ref{SIV}). 

On the other hand, NSs spin can be precisely determined either from pulsar radio or X-ray timing observations. A theoretical NS maximum spin frequency is given by the Kepler limit. Thus, spin measurements might also be used to constrain the EoS. Unfortunately, the maximum observed spin frequency of 716~Hz for a binary radio pulsar in Terzan 5 \cite{Hessels:2006ze} is not enough to stablish strong constraints on the NSs EoS \cite{Haensel:2009wa}, but this situation would change if a faster-spinning NS is discovered.

In other eclipsing binary systems containing pulsars with low-mass stellar companions, higher masses have been estimated, but with significantly larger systematics errors. In these systems, pulsars strongly irradiate their stellar companions, leading to less reliable mass determinations. For instance, a mass of $M = 2.40 \pm 0.12$~M$_\odot$ has been estimated
for PSR B1757+20 \cite{vanKerkwijk:2010mt} and, more recently, a mass of $M = 2.27 \pm 0.17$~M$_\odot$
for PSR J2215+5135 \cite{linares2018}. However, uncertainties in the inclination, make
the NS mass also compatible with lower values. Independent measurements of the orbital
inclination of these systems could lead to more precise mass determinations \cite{Romani:2015gaa}.

Although maximum-mass measurements alone can provide constraints on the nucleon interactions \cite{Hebeler:2013nza}, the current limits at $\sim2$~M$_\odot$ do not rule out the appearance of exotic particles in the core of NSs \cite{Oertel:2014qza,Fortin:2017cvt}. Radius determinations are needed to provide stronger constraints on their EoS. However, a very precise simultaneous determination of the mass and radius of the same NS is still pending. Several approaches based on X-ray observations were developed for this purpose, from the study of thermal emission of isolated NSs to accreting NSs in quiescent state and thermonuclear X-ray bursts in low-mass XRBs. In general, radii observations are challenging and remain strongly model dependent. Measurements are affected by external uncertainties on the interstellar extinction and source distance, but also by intrinsic causes as anisotropies in the emission, magnetic fields, rotation and specially atmosphere composition and residual accretion in XRBs (see Ref. \cite{Miller:2016pom} for a thorough analysis of the different issues). Depending on the type of system, different (dis)advantages are found.

\subsection{Isolated NSs}

Isolated NSs, by definition, do not experience accretion. The X-ray thermal emission coming from these sources originates in their internal cooling. In these cases, radius measurements are extremely difficult due to distance indetermination and poorly known atmosphere compositions. Moreover, in these sources it is quite hard to properly estimate their masses. NSs are born in violent supernovae explosions with very high initial temperatures. Isolated NSs experience long-term cooling throughout their lives. This cooling directly depends on their internal composition, being mainly governed by neutrino emissivities and heat-transport coefficients. If present, superfluidity can strongly modify the cooling characteristic timescales (see Ref. \cite{Potekhin:2015qsa} for a detailed review). NS cooling calculations present many uncertainties due to several not-fully-understood processes, but the observation of an enhanced cooling for about ten years in the young NS in Cas~A \cite{Heinke:2010cr}, may give direct information on its internal composition. Despite this result having been questioned \cite{Posselt:2013xva,Posselt:2018xaf}, if this rapid cooling is real, it could be explained by a transition to a neutron superfluidity phase in the NS core \cite{Page:2010aw}, which would become the first evidence of superfluidity and superconductivity at supra-nuclear densities within a NS. In a recent paper,  the authors of Ref. \cite{beloin2018} considered the cooling of several isolated NSs and found that, for a nuclear EoS, the threshold density for direct Urca processes is reached in NSs with masses in the 1.7--2.0~M$_\odot$ range. However, 2~M$_\odot$ NSs are less likely to cool down by the direct Urca process due to neutron-triplet superfluidity suppression.

\subsection{Quiescent emission from NS cooling}

NSs in low-mass XRBs tend to be discovered by wide-field X-ray monitors during transient episodes of strong accretion ({\it i.e.} the outburst state). In those episodes, their luminosity is highly increased and the NS crust significantly heated. When accretion stops, or strongly reduces, the NS enters in the so-called {\em quiescent} state. During quiescence, the emission is dominated by thermal radiation originating at the NS surface, which can be studied using pointed X-ray telescopes like {\em Chandra}, {\em XMM-Newton} and {\em NuSTAR}. If the distance to the source is known, a direct estimation of the NS radius can be achieved by fitting a NS atmosphere model to the time-evolving spectra. By means of a systematic study of five quiescent sources, it was shown in Ref. \cite{Lattimer:2013hma} that, depending on the assumptions made on the NS atmosphere composition, the EoS of dense matter is either incompatible with a strong phase transition, or this possibility is allowed. They also predict that the most probable radius for a canonical NS of 1.4~M$_\odot$ should be in the 10.5--12.7~km range. Discrepancies arising from chemical composition uncertainties are well illustrated by two studies on NGC 6397: for a hydrogen atmosphere, the authors of Ref. \cite{Guillot:2014lla} estimated $R = 9.4 \pm 1.2$~km, while for helium, a radius larger than 2~km was found \cite{Heinke:2014xaa}. More recently, the authors of Ref. \cite{Ozel:2015fia} added several corrections like those resulting from the inclusion of NS rotation, and, based on a comprehensive analysis of 12 sources, they estimate the radius of a 1.5~M$_\odot$ NS to be in the range 10.1--11.1~Km.

\subsection{Thermonuclear X-ray bursts in accreting binary systems}

NSs in low-mass XRBs show episodes of unstable thermonuclear burning of fresh hydrogen and helium deposited on their surfaces by the accretion processes \cite{strohbild2006,Galloway:2006eq}, leading to bright flares called Type I X-ray bursts. Some of these bursts are strong enough to reach the Eddington luminosity. In those cases, the radiation pressure triggers the expansion of the outer layers of the NS atmosphere, leading to a Photospheric Radius Expansion (PRE) burst. During these events, the luminosity increases by a factor of $\sim$10 in about 0.5--5~s. After that quick rise, the flux decreases exponentially for 10--100~s down to the persistent level. After such fast expansion, the atmosphere contracts again until it reaches back the NS surface, at {\em touchdown}, starting the so-called cooling phase. If the cooling process is homogeneous throughout the NS surface, the radius of the NS can be estimated by fitting the time-resolved spectra during the cooling phase of the bursts. Moreover, assuming that the Eddington luminosity is reached at the touchdown point, the mass and the distance to the NS can also be estimated if the chemical composition is known \cite{vanparadijs1979}.

Although the method seems very powerful, the strength of the attainable constraints is limited by several uncertainties associated with some assumptions: distance and atmospheric composition have a strong impact on the results \cite{strohbild2006,Lattimer:2013hma}, and it is not fully clear that touchdown is completely reached in all PRE bursts neither if that  the cooling phase strictly follows the expected behavior \cite{Steiner:2010fz,Garcia:2013ze}. Moreover, the presence of residual accretion affecting the spectral fits cannot be fully discarded. Following this approach and considering a carefully selected set of sources, the authors of Ref. \cite{Ozel:2016oaf} found that the radius of an 1.4~M$_\odot$ star is favored in the 9.8--11.0~km range. However, despite statistically significant constraints are often found using this method, results may suffer of systematics that should be evaluated with caution \cite{Miller:2016pom}.

A specific spectral model for the cooling phase of NS X-ray bursts was developed to constrain the mass and the radius of the NS in 4U 1724--307 assuming different chemical compositions \cite{Suleimanov:2010bp}. The method was applied later to three sources showing X-ray bursts at low mass-accretion rates, leading to radii estimations in the 10.5--12.8~km range for a canonical 1.4~M$_\odot$ NS \cite{Nattila:2015jra}. More recently, the authors of Ref. \cite{Nattila:2017wtj} fitted the same spectra using atmospheric burst models and obtained preferred radii and masses of 12.4$\pm$0.4~km and 1.9$\pm$0.3~M$_\odot$, respectively. This method allows to simultaneously fit constraints to the atmosphere composition.

\subsection{Emission lines}

The detection of an atomic-line feature at the surface of a NS would serve to directly estimate its gravitational redshift, and thus the ratio between it mass and it radius. The authors of Ref. \cite{Cottam:2002cu} have reported a narrow absorption feature in EXO~0748--676, but more recent observations failed to confirm the detection. The spinning of NSs should widen emission lines, making them much harder to detect, but recent studies suggest that line profiles from rotating NSs might actually be narrower than the initially predicted \cite{Baubock:2012bj}.

\subsection{Glitch activity}

In two recent works the glitch population has been used to probe the masses of NSs. The standard glitch scenario, assuming that the entire excess of angular momentum in the core is transferred to the crust during the largest glitch detected in a given pulsar, the authors of Ref. \cite{Pizzochero:2016ccv} placed limits on the NS mass. For example, for Vela and PSR J0537--6910, they found masses of 1.35$\pm$0.08~M$_\odot$ and 1.25$\pm$0.06~M$_\odot$, respectively. A different approach was suggested by the authors of Ref. \cite{Ho:2017ipg}, who performed a detailed thermal evolution model using a nuclear EoS with superfluidity. For the same two pulsars, the later authors obtained 1.51$\pm$0.04~M$_\odot$ and 1.83$\pm$0.04~M$_\odot$.

In turn, the highly-magnetized AXP 1E 2259+586 has experienced a so-called {\em anti-glitch}, or a sudden spin-down \cite{Archibald:2013kla}. This glitch was accompanied by an X-ray burst. As its behavior cannot be accounted by the standard glitch theory, different models assuming an internal \cite{2013Natur.497..574D,Kantor:2014dia,Garcia:2014moa} or external \cite{Tong:2013oct} origin have been proposed. More recently, new anti-glitches have been reported in the accreting pulsar NGC 300 ULX-1 \cite{paul2018}. Despite the present uncertainties, a precise detection of a glitch in a pulsar with known mass in a binary system would substantially increase our understanding of NS interiors.

\section{Post GW170817 era and restrictions to the NS equation of state}
\label{SIV}

Every time a new electromagnetic (EM) window has been  opened a revolution has occurred. Seeing in infrared, radio, x-rays and $\gamma$-rays has decisively changed
our understanding of the Universe. Three years ago a completely different window has opened up: we started to see in gravitational waves. This new ability allows us to study the most energetic events in the Universe from a radically different point of view and  to detect objects we had not even imagined. Gravitational
waves are not absorbed or reflected by matter like electromagnetic radiation,
for this reason through their detection it is possible to see objects without the EM problems 
Astronomy has. Moreover, as the Universe become transparent to GWs much earlier
than to EM waves, their detection and study would help us to understand the first
instants after the Big Bang.

Until 2015, GWs had been only indirectly detected and the most
important of these indirect observations was associated to the Hulse-Taylor
pulsar \cite{hulsetaylor1974,taylor1982,Weisberg2016}. On September 14th, 2015 at 9:50:45 UTC, the LIGO detectors in Hanford,
Washington and Livingston, Louisiana, detected GWs emitted during a binary
black hole merger \cite{GW1}. This detection marked the beginning of a new era: the
gravitational-wave astronomy era. Analysis of the signal of GW150914 that lasted
$\sim 0.2$s has confirmed the existence of both gravitational waves
predicted by Einstein and of solar-mass black holes. 
For
the first time in history, General Relativity was tested in the
strong-field regime, and no deviation from theoretical predictions was found \cite{abbott2016PhRvL.116v1101A}.

After decades of development, LIGO and Virgo collaborations had achieved the direct detection of GWs emitted from ten binary black hole mergers: GW150914, GW151012, GW151226, GW170104, GW170608, GW170814, GW170729, GW170809, GW170818 and GW170823 (see Ref. \cite{GWTC-1} and references therein). By getting some adjustments and improvements in the detectors sensitivity, the strongest evidence to confirm the hypothesis that mergers of binary stars are the cause of short gamma-ray bursts would appear in 2017: the detection of the GW170817 and GRB 170817A counterpart from a merger of two NSs.

\subsection{The GW170817  and  GRB 170817A connection}

The LIGO/Virgo Collaboration detected for the first time GWs emitted from a
binary NS merger GW170817 \cite{GW5}. On August 17, 2017, at 12: 41: 04.4 UTC the fusion of the two NSs was registered. The first electromagnetic signal detected was GRB 170817A, a short gamma-ray burst, detected by Fermi and INTEGRAL 1.74 $\pm$ 0.05 seconds after the fusion time and which lasted approximately 2 seconds. The compact objects that merged have been
determined to be compatible with low-mass NSs with masses in the range $\sim 1.2$ to $\sim
1.6M_\odot$. The nature of the $\sim 2.7M_\odot$ remnant is not clear; it could
be either an extremely massive NS or a low-mass BH \cite{pooley2018}.

The combined data obtained from GW170817 and GRB 170817A unambiguously
associates binary NS mergers with short gamma-ray bursts (sGRB) and helped us to
understand the nature and properties of these systems, constituting central engines of sGRB
\cite{2018arXiv180207328V}. 
The NS-NS merger took place at a
distance of $\sim 40$Mpc form the Earth.  The analysis of the
data from GW170817 and GRB 170817A sheds some light on some fundamental physical
aspects. It allows us to constrain the difference between the speed of light in
vacuum, $c$, and the speed of gravity, $c_g$, a difference determined to be
between $-3 \times 10^{-15} < (1-c_g/c) <7 \times  10^{-16}$ \cite{2017ApJ...848L..13A}.
Moreover, it served to put limits on the violation of Lorentz invariance and
test the equivalence principle.

\subsection{Mergers of NSs and restriction to the EoS from GWs}
\label{mergers}

In a NS-NS merger simulation , linearized gravity can not be used to describe the system properly. In such events, the dynamical evolution is critically influenced by the backreaction of the GWs emitted. This kind of situation is studied in the ``post-Newtonian'' (PN) approximation. Within this scenario, backreaction is considered when the 2.5PN order is analyzed (for details, see Ref. \cite{blanchet2014}). PN expansions do not converge when the velocities involved are a considerable fraction of the speed of light in vacuum. To overcome this problem different proposals of resummation of series have been developed. One commonly used in the literature is the so called Effective One Body model, first developed for binary BH mergers \cite{buonanno1999} and extended later to calculate waveforms produced in binary NS mergers using full 3D numerical relativity \cite{damour2010}.  Simulations solving the dynamics of mergers in full 3D general relativity spread rapidly also thanks to free access codes like {\texttt{Einstein Toolkit}} \cite{einsteintoolkit2012} and {\texttt{LORENE}} (see, \url{https://lorene.obspm.fr/}). Final stages of the merger and the post-merger phase must be studied using numerical relativity. A review of the initial data of binary NSs mergers and advanced simulations with approximate treatments of gravity can be found in Ref. \cite{Baiotti_2017}.

Binary NS merger simulations
in full 3D general relativity were performed for the first time in the early 2000's \cite{Shibata:1999wm,Shibata:2002jb}. In these pioneering works, mergers of polytropic NSs with equal masses were studied. Since then, huge  advances have been made 
in the area of numerical modeling of binary NS mergers. Many different EoS for NS in merger simulations have been used and a range of mass and spins of the components of the binary system has been covered to produce a comprehensive catalogue of waveforms that can be compared with observable data (for details, see Ref. \cite{Dietrich:2018phi} and references therein).

GW waveforms produced during the early stages of a NS merger are characterized
almost unambiguously by the tidal deformability $\Lambda$, which is a measure of
how a NS is deformed due to the gravitational field of the companion compact object
(see, for example, Ref. \cite{flanagan2008}). A qualitative study of the
astrophysical implications of measuring the pre-merger waveform in NS-NS or
NS-BH mergers is performed in Ref. \cite{read2009}. In this paper, the authors
conclude that, as waveforms deviate from the pointlike particle approximation,
this information would be a powerful tool to help unravel the nature of compact
objects EoS, helping us understand the behavior of matter when it is compress
to extreme NS-like pressures.

Using linearized theory, the tidal deformability can be written in terms of the
dimensionless tidal Love number $k_2$ (see, Ref. \cite{hinderer2010}, and
references therein). Recently, the authors of Ref. \cite{2018arXiv181010967H} have performed calculations using hybrid EoS within the ``Constant-Speed-of-Sound'' (CSS) parametrization, a generic parametrization of the quark matter
EoS with a density-independent speed of sound \cite{cssoriginal}, to show how tidal deformability behaves in the presence of sharp phase transitions.

\begin{figure}[!htbp]
\centering
\includegraphics[width=0.7\textwidth]{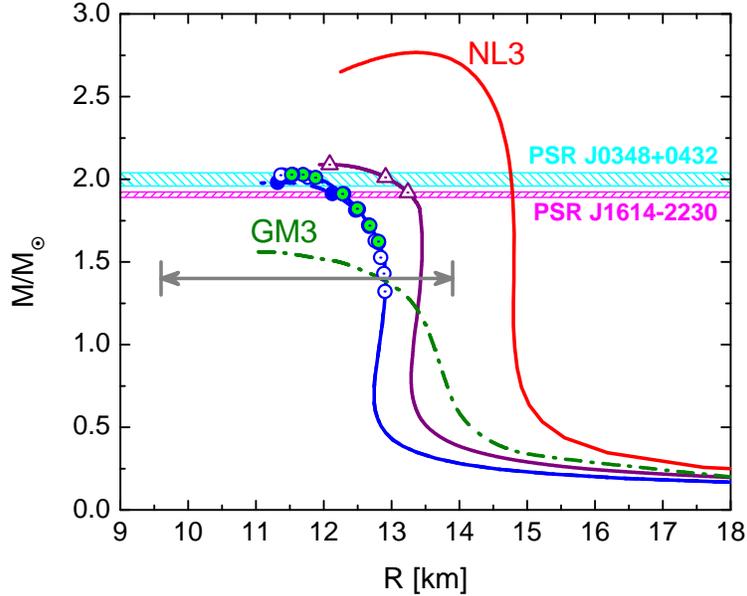}
\caption{(Color online) M-R relationship for different EoS. Lines with triangles and circles indicates HSs with color superconducting cores. The gray arrow corresponds to the  radius constraint from $GW170817$ shown in figure \ref{canonical_radius} . }
\label{masa_radio}
\end{figure}

\begin{figure*}[htb]
\centering
\includegraphics[width=0.7\textwidth]{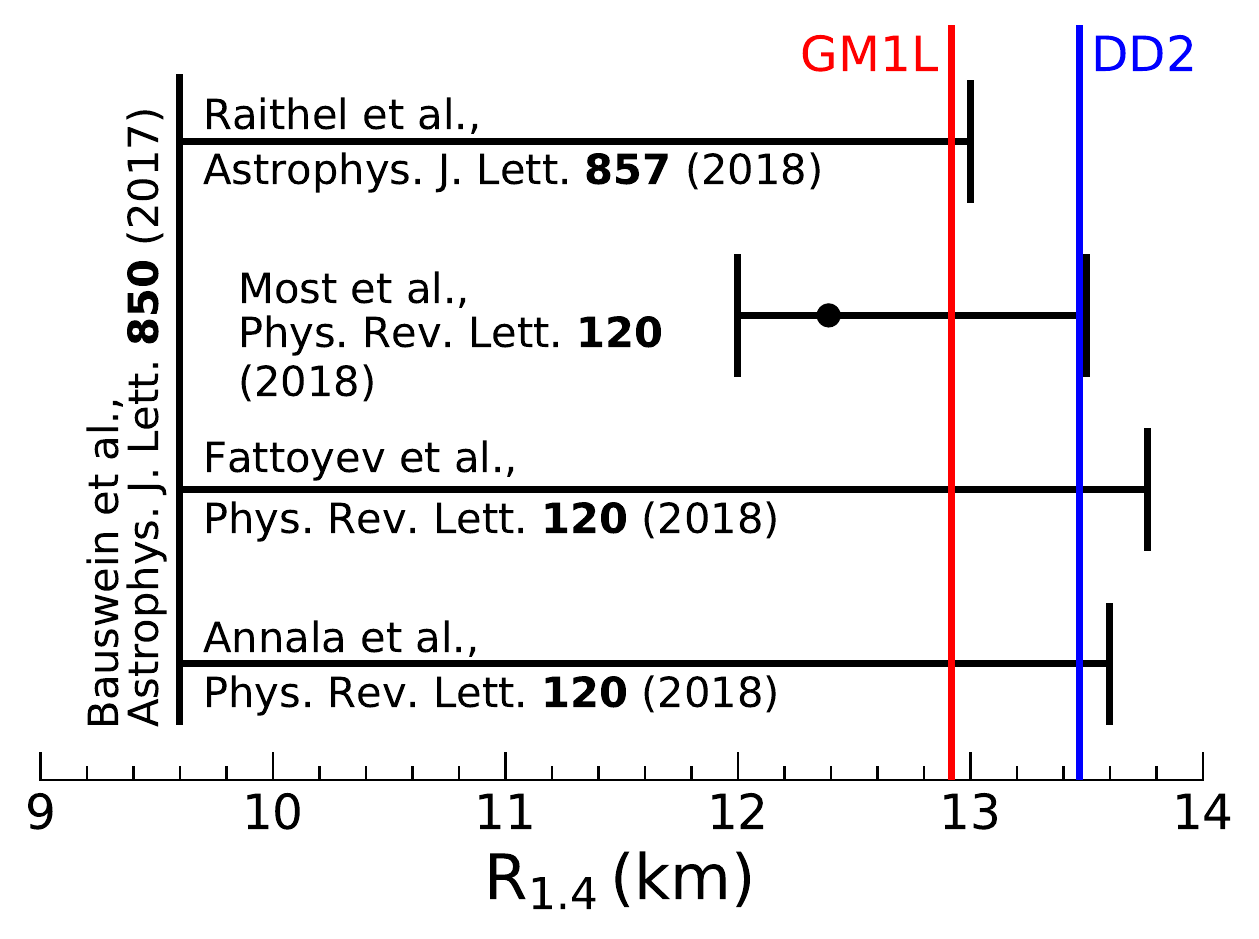}
\caption{(Color online) Radius constraints derived from  $GW 170817$ data, which are reproduced by the DDRMF model EoSs
  discussed in this review.}
\label{canonical_radius}
\end{figure*}

Analysis of data from GW170817 put an upper limit on the effective tidal
deformability $\tilde{\Lambda}$ ({\it i.e.} the mass weighted-average of the
individual tidal deformabilities) as well as the chirp mass of the NS-NS binary
system.
The $\tilde{\Lambda} < 800$ limit allowed to conclude, with
$90\%$ confidence, that the radius of a NS with a mass of $1.4M_\odot$, $R_{1.4}$,
cannot exceed $\sim 13.6$ km \cite{raithel2018}. Other studies (see, for
example, Refs. \cite{Annala:2017llu,most2018,malik2018,2018PhRvL.120q2702F,Bauswein:2019ybt}) arrive to similar
conclusions. This result can be used to discard several hadronic EoS, like the
famous NL3 and GM3 parametrizations of the Walecka  RMF as 
shown in figure \ref{masa_radio}. In addition, it has been argued that the
radius of the non-rotating maximum-mass configuration must be larger than $\sim
9.6$ km \cite{Bauswein:2017vtn}. The improved NS EoSs obtained with DDRMF
parametrizations are compatible with the constraint of $R_{1.4}$ derived from
the GW170817 data (for details, see figure \ref{canonical_radius}).
%here fig 7 to be fig 6

After a detailed analysis of data from GW170817 and its EM counterpart,
additional restrictions to the EoS have been found. According to Ref.
\cite{PhysRevLett.121.161101}, the pressure at two times saturation density must be
in the range $P(2\,n_0) = 3.5^{+2.7}_{-1.7} \times 10^{34}{\rm dyn/cm}^2$, and
at six times saturation density in the range $P(6\,n_0) = 9.0^{+7.9}_{-2.6}
\times 10^{35}{\rm dyn/cm}^2$ at the $90\%$ confidence level. The first
restriction is not very stringent, but the second one serves to discard very soft
hadronic EoSs.
%here fig 6 to be 7

Multiwavelength (X-ray, UV, optical, infrared, and radio bands) analyses of
the EM counterpart of GW170817 (called AT2017gfo) tentatively constraint
$\tilde{\Lambda} > 400$ \cite{2018ApJ...852L..29R}. These constraints serve to
exclude hadronic EoSs that are either too stiff or too soft. Some criticisms have been raised up regarding the results presented in Ref. \cite{2018ApJ...852L..29R}. The authors of Ref. \cite{2018PhRvC..98d5804T} suggest that $\tilde{\Lambda} > 400$ restriction might have been an overestimation. 
The authors of Ref. \cite{2018arXiv181006109D} have studied the role of vector-isovector meson
cross interactions and their effect on a series of hadronic EoSs and on macroscopic
quantities, in particular the M-R relationship and the tidal deformability.
As a general result, they found the additional interaction reduces the radius of the
canonical NS, allowing all the presented models to satisfy the restrictions
imposed by GW170817.

Several model-dependent studies have been performed analyzing the existence of HS.
Pure quark matter cores in HS were analyzed using the Maxwell construction for
the hadron-quark phase transition in Ref. \cite{2018PhRvD..97h4038P}
Alternatively, several works indicate that HS might
exist but without pure quark matter in their cores, rather a hadron-quark
mixed-phase core
(see, for example, Refs. \cite{2018ApJ...857...12N,2018PhRvD..98h3013L}). Moreover,
the implications of the radial constraints on a $1.4M_\odot$ NS suggest that
HS with quark cores can exist only if the hadron-quark phase transition occurs
at an energy density similar to that of nuclear saturation \cite{2018arXiv180604763G}.

Another interesting series of works are those dedicated to finding GWs from the post-merger phase. In Ref. \cite{Clark:2015zxa} the relevance of the information of detecting GWs from the post-merger phase is emphasized, specially the connection of $f_{peak}$, the dominant post-merger frequency which contains information about the structure and composition of the compact remnant \cite{PhysRevD.86.063001,Torres-Rivas:2018svp}. In addition, it has been found that a strong discontinuity in $f_{peak}$  as a  function of the total mass would be an indicator of the coexistence of two families of compact stars: hadronic and quark matter stars \cite{Bauswein:2015vxa}. On the other hand, the authors of Refs.\cite{PhysRevLett.113.091104, PhysRevD.91.064001} have analyzed a large number of accurate numerical-relativity simulations of binaries with nuclear EoSs finding an universal correlation between the low-frequency peak related to the merger process and the total compactness of the stars in the binary system. This also provides a powerful tool to set constraints on the EoS. 

A search of GWs emitted in the post-merger phase of GW170817 has been performed, but no evidence of emission has been found yet \cite{Abbott:2017dke}. 

\section{GWs from isolated NSs}
    \label{SV}
        
The most promising astrophysical scenario in which isolated NSs could radiate gravitational waves is probably related to the birth of proto-NSs in core-collapse supernovae due to the rate of events and the energy involved  (see Ref. \cite{2018ApJ...861...10M} and references therein). Additionally, isolated NSs could radiate GWs if the stars are deformed and rotating. Other circumstances that could lead to emission of GWs from isolated NSs are pulsar glitches, magneto-elastic oscillations in magnetars, and “quasi-normal” mode oscillations excited by the rearrangement of a compact star after a hadron-quark phase transition or a star quake.

It is expected that future LIGO/Virgo observational runs might be able to detect not only GWs emitted during mergers of compact objects but also those emitted by isolated NS.

The EoS of NS matter could be constrained by the analysis of the GW emission of
stellar pulsation modes. Such oscillations could be excited in newly born
proto-NSs \cite{Camelio:2017nka} due to starquakes and glitches
\cite{Warszawski:2012zq}, by accretion in a binary system, or by the
rearrangement of the star following the conversion of a hadronic star into a
quark star \cite{Lugones:2002vj,Abdikamalov:2008df}. It can also occur in the
post-merger phase of a binary NS system if the remnant does not collapse into
a black hole \cite{Bauswein:2015vxa,Andersson:2009yt,Faber:2012rw}.

In this Section we present some basic aspects of NSs oscillations and their link to GW emission from isolated NSs. The focus will be put on non-radial oscillations, in particular on the $g-$mode whose detection could be a strong indicator of the occurrence of a sharp phase transition from hadronic matter to quark matter in the interior of NSs. We also briefly discuss the role of dynamical and secular instabilities  in NSs. 

\subsection{ Quasi-normal modes of NSs}

In a pulsating star, energy is lost by the emission of gravitational radiation;
such vibrations are not described by normal modes but by  ``quasi-normal''
modes. The pulsation equations are obtained by introducing small perturbations
in the Einstein equations \cite{NS-oscillations,Andersson:2009yt}.  A general linear
perturbation of a scalar quantity (e.g. $\delta \epsilon$ in the energy
density or $\delta p$ in the pressure) can be written as a sum of
quasi-normal modes that are characterized by the indices $(\ell, m)$ of the
spherical harmonic functions $Y_\ell^m$ and a time dependence of the form $e^{i
\omega_{\ell} t}$, where $\omega_l$ is a complex frequency (the real part
represents the actual oscillation frequency and the imaginary part the
damping produced by the emission of GWs). The perturbations in the fluid's
four-velocity $\delta u_{\mu}$ can be expressed in terms of vector harmonics,
and the metric perturbation $\delta g_{\mu \nu}$ in terms of spherical,
vector, and tensor harmonics. These perturbations can be classified as polar
or axial, according to their parity (which is defined as the change in sign
under a combination of a reflection in the equatorial plane and rotation by
$\pi$). Polar perturbations have parity $(-1)^{\ell}$ and axial perturbations
have parity $(-1)^{\ell+1}$. The important point here is that, for non-rotating
relativistic stars, it can be shown that the polar and axial perturbations are
completely decoupled (see Ref. \cite{Stergioulas:2003yp} and references therein).
Polar modes produce spheroidal fluid deformations, and axial modes toroidal
deformations.  Modes are said to be fluid if they involve a significant fluid
motion, and spacetime if they only involve vibrations of spacetime. Polar fluid
modes are analogous to the Newtonian fluid pulsations. The most well known in
the literature are the fundamental ($f$), the pressure ($p$), and the gravity
($g$) modes, but there are other relevant modes such as the shear mode in a
solid crust or the Alfv\'en modes in the presence of magnetic fields. In the
case of axial modes, there are torsional modes and rotational modes (Rossby and
inertial modes). There are also polar and axial spacetime modes, which only
involve vibrations of spacetime (the so called $w$ modes).

It is important to note at this point that the global properties of NSs, such as masses and radii, are
only marginally modified by rotation as long as the stars' spin
frequencies are well below the Kepler frequency.  With the exception
of pulsar J1748-244ad, which rotates at 716 Hz, this is the case for
all NSs known to date. Therefore, restricting ourselves to non-rotating compact objects, polar non-radial perturbations
introduce changes in the Schwarszchild line element that can be written as
\begin{eqnarray}
{\rm d}s^2 &=& -{\rm e}^{\nu (r)}\left(1 + r^\ell H_0Y_\ell^m {\rm e}^{i \omega t}\right){\rm d}t^2 - 2i\omega r^{\ell +1}H_1Y_\ell^m{\rm e}^{i \omega t}{\rm d}t{\rm d}r \nonumber \\
&&+ {\rm e}^{\lambda (r)}\left(1 - r^\ell H_0Y_\ell^m {\rm e}^{i \omega t}\right){\rm d}r^2 \nonumber \\
& & + r^2\left(1 - r^\ell K Y_\ell^m {\rm e}^{i \omega t}\right)\left({\rm d}\theta^2 + \sin ^2 \theta {\rm d}\phi^2 \right). 
\end{eqnarray}
The Lagrangian displacements that perturb the fluid can be written as
\begin{eqnarray}
\xi^r &=& r^{\ell -1}{\rm e}^{-\lambda (r)/2} W(r)Y_\ell^m {\rm e}^{i \omega t} , \nonumber \\
\xi^\theta &=& - r^{\ell -2}V(r)\partial_\theta Y_\ell^m {\rm e}^{i \omega t},\\
\xi^\phi &=& -\frac{r^\ell}{(r\sin \theta)^2} V(r) \partial _\phi Y_\ell^m {\rm e}^{i \omega t}. \nonumber
\end{eqnarray}
The functions $H_1$, $W$, $K$ and $X$ (a linear combination of $V$, $W$ and
$H_0$) satisfy a system of first order differential equations. Outside the
compact object, fluid perturbations vanish and the perturbations are ruled by
the Zerilli equation. For a given $\ell$, the quasi-normal modes of the star are
solutions of the pulsation equations that behave as pure outgoing waves at
infinity, are regular at the NS center, and verify that the interior solution
matches continuously with the exterior perturbation at the NS surface (see Ref.
\cite{vf-lugones2018} and references therein).

\subsection{Exotic stars, HSs and oscillation modes}
Stellar pulsation modes may emit
gravitational radiation that could be detected by the LIGO/Virgo observatories
as well as by a new generation of planned GW detectors such as KAGRA, the Einstein
Telescope and IndIGO (see, for example, Ref. \cite{gw-pospects2018} and references
therein). Particularly, the identification of features in the $f$ and $g$ modes may signal 
unequivocally the appearance of deconfined quark matter in NSs or reveal the existence of exotic stars.

\begin{figure}[!htbp]
\centering
\includegraphics[width=0.60\textwidth]{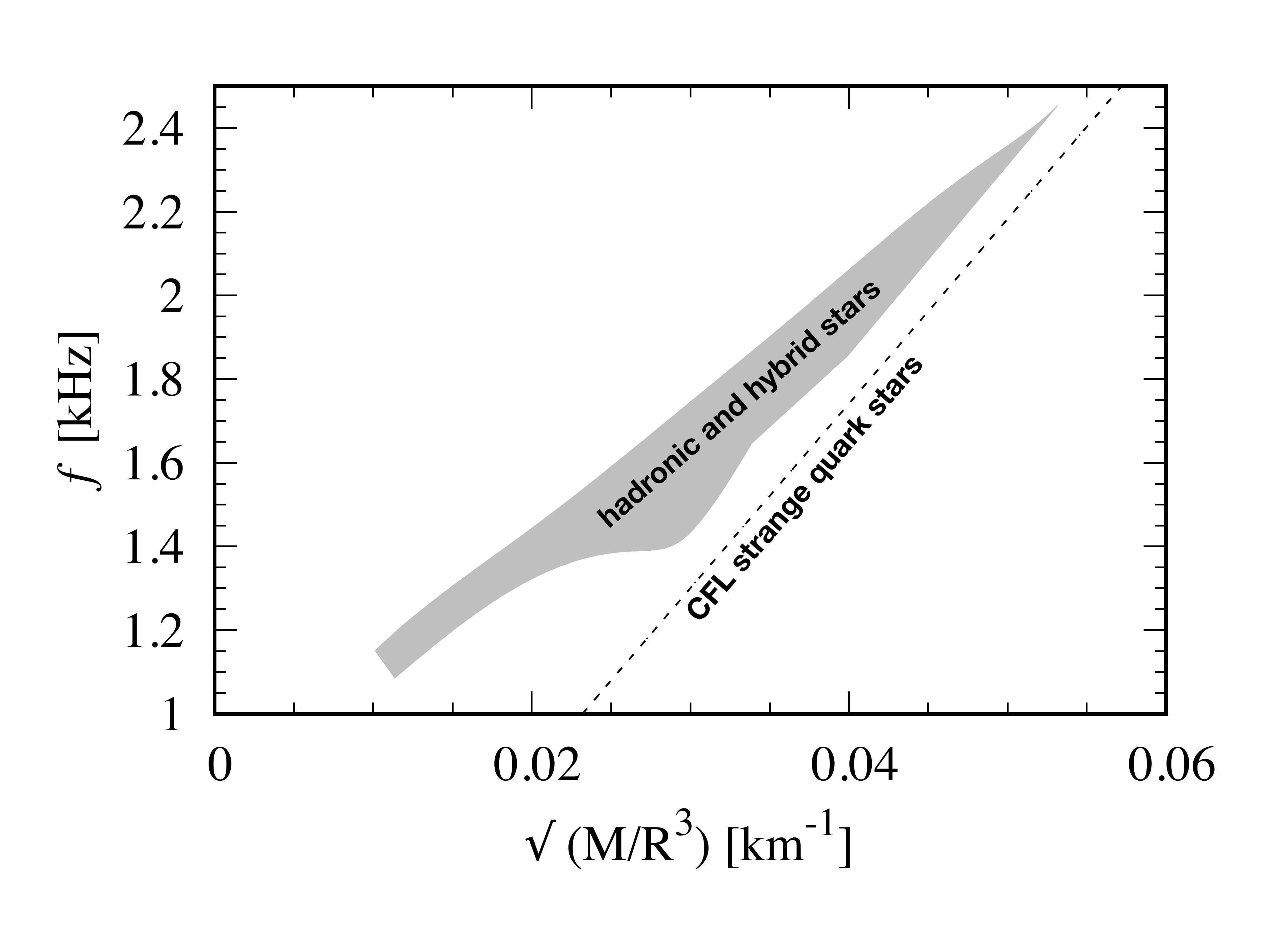}
\caption{(Color online) The frequency of the $f$-mode as a function of the
  square root of the average density of a NS. Hadronic and HSs fall within a
  rather model independent shaded region which was obtained using a set of EoSs
  that interpolate between reliable calculations for low-density nuclear matter
  and high-density quark matter (see Ref. \cite{vf-lugones2018} for details). For
  comparison, we also show a universal relation that parametrizes the results
  for color-flavor locked strange quark stars \cite{Flores:2017hpb}.}
\label{fig:fmode}
\end{figure}

The energy of the fundamental fluid mode ($f$) is
quite effectively channeled into GW emission. Its properties have been
extensively analyzed for several EoSs. An interesting and
observationally useful property of the $f$-mode is that it can be described by
universal fitting formula that are argued to be quite EoS-independent. In fact,
it has been shown that the oscillation frequency $f$ of the fundamental mode
has a reasonably linear dependence on the square root of the average density,
and the damping time $\tau$ can be fit with a simple formula involving the
stellar mass and radius
\cite{Andersson:1997rn,Benhar:2004xg,Chirenti:2015dda,Lau:2009bu}. These
results suggest that the mass and radius of a compact object could be inferred
if $f$ and $\tau$ were detected by the new generation of gravitational wave
detectors. In a recent work, the properties of the fundamental mode have been
systematically studied using a set of EoS connecting state-of-the-art
calculations at low and high densities \cite{vf-lugones2018}. Specifically, a
low density model based on the chiral effective field theory (EFT) and high
density results based on perturbative Quantum Chromodynamics (QCD) are matched
through different interpolating polytropes fulfilling thermodynamic stability
and subluminality of the speed of sound, together with the additional
requirement that the equations of state support a two solar mass NS. Using
this family of EoSs, the frequency and the damping time of the $f$-mode are
constrained within narrow, quite model-independent windows (see figure
\ref{fig:fmode}) which are in good agreement with many previous works that used
a variety of phenomenological EoSs. In principle, GWs arising from the $f$-mode
of \textit{non-rotating} compact stars should fall within the window presented
in figure \ref{fig:fmode}. If not, the emitting object could be some kind of
exotic star (see, for example, Ref. \cite{Flores:2017hpb}).

The previous results hold for non-rotating objects. However, it is expected that all NSs should have at least some degree of rotation. Moreover, promising scenarios for detecting the $f-$mode such as NS mergers are likely to involve rapidly spinning NSs. In spite of its relevance, a generalization of GW asteroseismology to rotating objects has only been carried out recently within the Cowling approximation (where the spacetime metric is unperturbed and only the fluid perturbations are considered). The $f$-mode frequency as measured in the stellar rotating frame can be written as a quadratic polynomial in terms of  the angular frequency $\Omega$  \cite{PhysRevD.83.064031,PhysRevD.88.044052}. For objects spinning at a significant fraction of the mass shedding angular frequency, the $f$-mode oscillation frequency is very different from the non-rotating value. Moreover, when rotation is sufficiently fast, NSs can be destabilized by the so-called Chandrasekhar-Friedman-Schutz  instability \cite{PhysRevLett.24.611,Friedman:1978hf}, and some non-axisymmetric modes may grow in amplitude rather than decaying under the emission of GWs (see Section \ref{instabilities}).

Another important family of fluid ``quasi-normal'' modes are the gravity modes
($g$) which arise because gravity tends to smooth out material inhomogeneities
along equipotential level-surfaces. They appear only in the presence of a
significant temperature or composition gradient or as a consequence of sharp
discontinuities due to first order phase transitions
\cite{Andersson:2009yt,AnderssonKokkotas:1996,AndersonKokkotas:1998,vasquez-lugones:2014}.
Unlike pressure ($p$) modes where the radial displacement is much bigger than
the tangential one, for $g$-modes the tangential displacement is much bigger
than the radial one. 

\begin{figure}[!htbp]
\centering
\includegraphics[width=0.7\textwidth, angle=0]{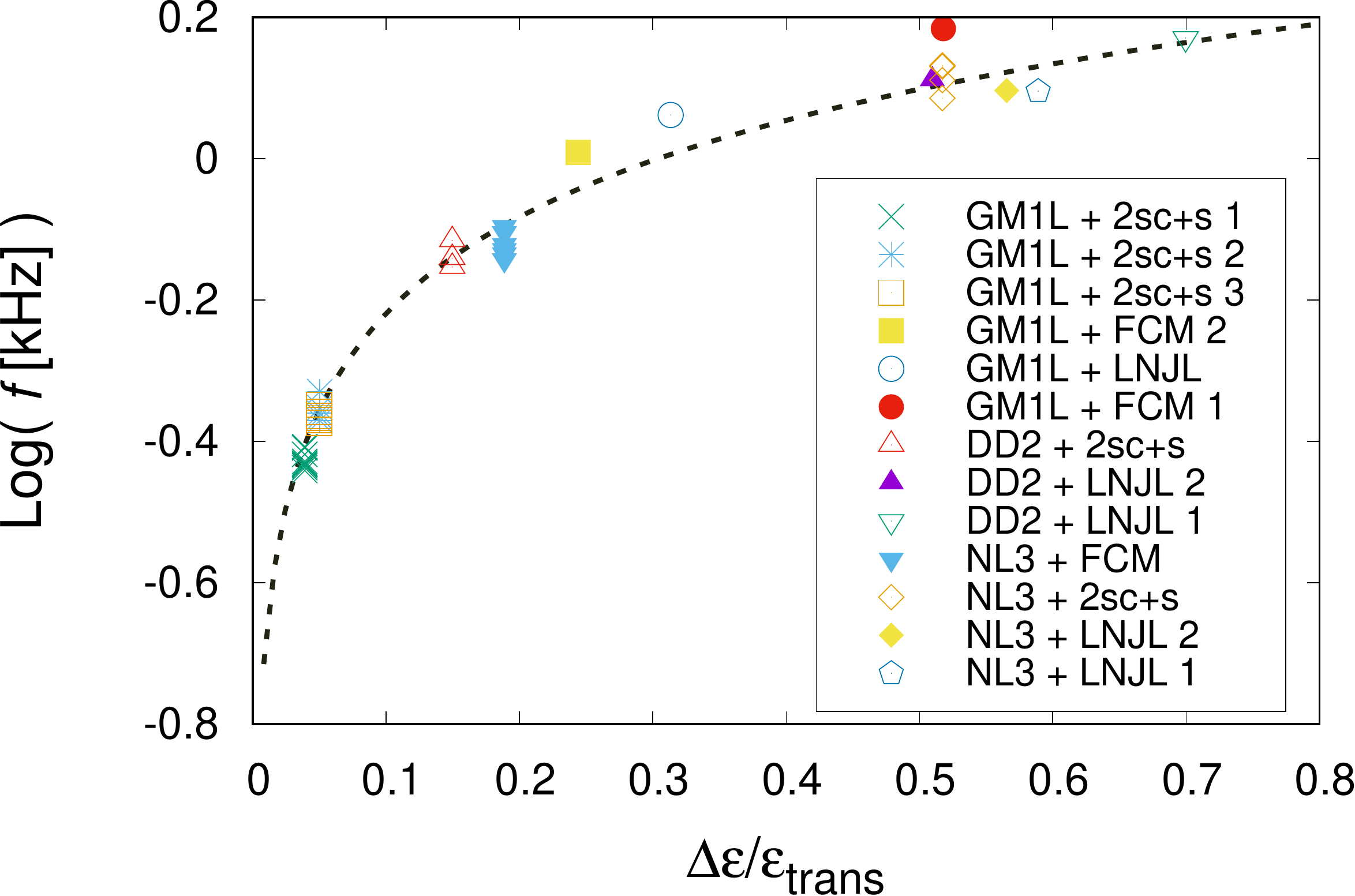}
\caption{(Color online) Universal relationship between the $g$-mode frequencies and the $\Delta \epsilon/\epsilon_{\rm trans}$ parameter of the CSS parametrization for quark matter, 
  obtained for different EoSs. Figure adapted from Ref. \cite{Ranea-Sandoval:2018bgu}.}
\label{css_universal}
\end{figure}

If density discontinuities actually exist inside NSs due to the existence of a
deconfined quark matter core, the associated $g$-mode would be a valuable
indicator of its presence. Moreover, as shown in Refs. \cite{Sotani:2001bb,Miniutti:2002bh,Sotani:2010mx,vasquez-lugones:2014,Ranea-Sandoval:2018bgu}
the $g$-mode frequency strongly depends on the amplitude of the discontinuity, as it is shown in figure \ref{css_universal}. 
Thus, it could be used to infer the value of the density jump and shed some
light on the properties of the hadron-quark transition.

NJL-like models typically yield equations of state that feature a first-order
transition to quark matter. These kinds of models have been extensively used to
describe the matter in the inner cores of HSs
\cite{Blaschke:2005uj,Orsaria:2012je,Orsaria:2013hna,Ranea-Sandoval:2015ldr,Ranea-Sandoval:2017ort}.
The first order deconfinement transition of hadronic matter at low temperature
but high density can be modeled through a hybrid EoS, which describes NS matter at low densities in terms of hadrons and of deconfined quarks at
high densities, assuming a sharp first-order phase transition between hadronic
and quark matter and considering the CSS parametrization.  Thus,
supposing the hadron-quark surface tension is high enough to disfavor mixed
phases, and restricting EoSs to those that allow NS to reach 2$M_{\odot}$, it was
found that the appearance of the quark matter core either destabilizes the star
immediately (this is typical for non-local NJL models) or leads to a very short
HS branch in the mass-radius relation (this is typical for local NJL models).
By using the CSS parametrization it has been shown that the reason for the
near-absence of HSs is that the transition pressure is fairly high and the
transition is strongly first-order \cite{Ranea-Sandoval:2017ort}. This
situation changes if the possibility of formation of a color superconducting
phase is considered in the description of dense matter. Color superconductivity
lowers the transition pressure, making it possible to construct stable
hydrid stars containing a diquark matter core. According to the results
presented in Ref. \cite{Ranea-Sandoval:2018bgu}, the maximum mass star corresponding
to the different families constructed through the hybrid EoS with diquarks
would be composed of 60-75 $\%$ quark matter in a superconducting phase.

\begin{figure}[!htbp]
\centering
\includegraphics[width=0.7\textwidth]{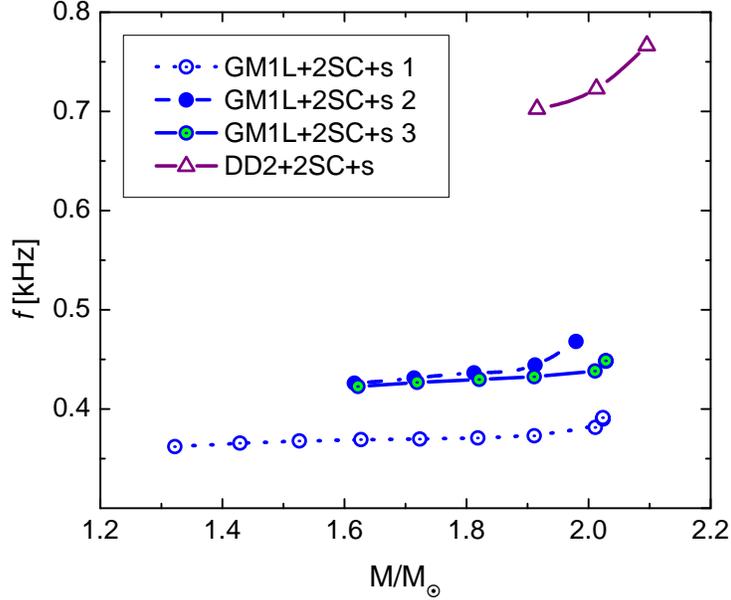}
\caption{(Color online) Frequencies of the g-mode for hybrid stars containing diquark matter cores \cite{Ranea-Sandoval:2018bgu}.}
\label{g_mode}
\end{figure}

In figure \ref{css_universal} the frequencies of the $g$-modes have been obtained for the different hybrid EoSs as a function of the quotient between the quark-hadron energy gap, $\Delta \epsilon$, and energy density at the phase transition, $\epsilon_{\rm trans}$, one of the parameters of the CSS parametrization. This result suggests that the obtained relationship is of {\it universal} nature and could be used to better our understanding of the behavior of the dense matter in NSs. After a detection of a $g$-mode, it may be possible to infer the existence of a sharp phase transition inside NSs and to deduce a value for one of the CSS parameters, constraining the EoS.

Figure \ref{g_mode} shows the frequencies of the $g$-modes as a function of gravitational mass for compact objects constructed using the hybrid EoSs presented in Ref. \cite{Ranea-Sandoval:2018bgu}.

As it has been mentioned in subsection \ref{mergers}, some of the features of the postmerger evolution could reveal important information about the properties of the EoS of compact objects. In particular,  the study of the postmerger phase could provide evidence of the existence of a sharp phase transition in the interiors of NSs. One of the key issues would be the presence of a discontinuity in the dominant post merger GW frequency, $f_{peak}$.  The imprint on GWs emitted in a NS merger due to a discontinuity in the EoS has also been analyzed in Ref. \cite{PhysRevLett.122.061102}. Using a particular hybrid, temperature-dependent EoS \cite{fischer2018}, the authors found that the frequency of the dominant post-merger mode suffers a significant modification when compared to the waveforms emitted when normal NSs merge.
Performing simulations using full general relativity, observational signatures of the
appearance of quark matter in the post-merger high-temperature phase were found.
When it occurs, this phase transition leads to a ringdown GW signal that can be
distinguished from the one produced if a purely hadronic EoS is used
\cite{PhysRevLett.122.061101}. In this way, the GWs emitted as a consequence of a binary
NS merger can be used to probe the behavior of matter at high-temperature and
density and to study, from a different point of view, properties of the QCD phase
diagram.

\subsection{Intensity of GW signals}

If the frequencies of the mentioned oscillation modes were detectable with the Advanced LIGO/Virgo detector or Einstein telescope, we could probe the internal composition of NSs. To set constraints on the different theoretical possibilities of EoSs, we need better observation data, which probably comes with the improvement in sensitivity of the third generation detectors.

From the data, and assuming certain parameters, for example that detection requires
certain signal-to-noise ratio and the distance to the source,  it is possible to estimate the minimum energy required in each mode to be detectable with Advanced LIGO/Virgo or Einstein telescope, considering the respective characteristic frequency. Useful order of magnitude calculations that allow us to estimate the amount of
energy that must be channeled in a particular GW mode in order to be detectable
with a given detector can be written \cite{Andersson:2009yt}

\begin{equation} \label{energy}
  \frac{E_{\rm gw}}{M_\odot c^2} = 3.47 \times 10^{36} \left(\frac{S}{N}\right)^2 \frac{1+4Q^2}{4Q^2} \left(\frac{D}{10{\rm kpc}}\right)^2\left(\frac{f}{1{\rm kHz}}\right)^2\left(\frac{S_n}{1{\rm Hz}^{-1}}\right)^2,
\end{equation}

\noindent where $S_n$ is the noise power spectral density of the detector, that
for a frequency of 1kHz is $S_n^{1/2} \sim 2 \times 10^{-23}{\rm Hz}^{-1/2}$
for Advanced LIGO/Virgo and is supposed to be, in the same band, $S_n^{1/2}
\sim 10^{-24}{\rm Hz}^{-1/2}$ for the Einstein telescope. $D$ is the distance
to the source, $f$ the frequency of the mode, $Q=\pi f \tau$ the quality
factor (which introduces a slight dependence with the damping time) and $S/N$
the desired signal-to-noise ratio. 

%%%%%%%%%%%%%     TABLE 1    %%%%%%%%%%%%%
\begin{table}[tbh] % \label{mode-energetics}
\centering
\begin{tabular}{c c c  c c}
\hline \hline
Detector &  Distance &     $f$-mode          &        $p_1$-mode              &   $g$-mode              \\
\hline
LIGO/Virgo &  10 kpc   &  $8.9\times 10^{-8}$   &    $3.2\times 10^{-6}$      & $8.0\times 10^{-9}$    \\
LIGO/Virgo &  15 Mpc   &  0.2   &       7.2           &  $1.8\times 10^{-2}$ \\
Einstein  &  10 kpc     &  $8.9\times 10^{-10}$     &  $7.2\times 10^{-8}$ &  $1.8\times 10^{-10}$  \\
Einstein  &  15 Mpc    &  $2\times 10^{-3}$    &   0.16  &  $4\times 10^{-4}$ \\
\hline \hline
\end{tabular}
\caption{The minimum estimated energy (in units of $M_{\odot} c^2)$ required in
  each mode in order to lead to a detection with signal-to-noise ratio of S/N =
  8 from a pulsating NS at different distances (local event or at the distance
  of the Virgo cluster). We use typical order of magnitude frequencies and
  damping times for each mode.}\label{mode-energetics}
\end{table}

In table \ref{mode-energetics} we present the minimum energy that should be
channeled into a specific oscillation mode in order to be detected. To obtain this estimate, we have used typical frequencies of 2 kHz, 7 kHz and 1 kHz for the excited modes $f$, $p_1$ and $g$, respectively \cite{vasquez-lugones:2014,Ranea-Sandoval:2018bgu}. The
obtained energetics are not unrealistic; numerical core-collapse simulations
estimate that the energy emitted in GWs in such an event should be of the order
of $10^{-9} - 10^{-8} {\rm M_{\odot} {c}^2}$ \cite{muller2004}. A conservative upper limit
might be taken from simulations of the collapse of a rapidly rotating NS to
form a black hole. Such an event would radiate GWs with a total energy of $\sim
10^{-6} {\rm M_{\odot} {c}^2}$ \cite{baiotti2006}.

\subsection{Other sources of GWs: dynamical and secular instabilities}
\label{instabilities}

The above considerations were focused on non rotating stars. However, rotating compact objects may be affected by  nonaxisymmetric  dynamical and secular instabilities. The most studied  type of rotational dynamical instability is the so called  bar-mode instability. The $m=2$ bar-mode instability  would be excited when $\beta=T/|W|$ is larger than 0.24 \cite{SBS2000,Saijo01}, where $T$ the rotational kinetic energy and $|W|$ the gravitational energy. However, differential rotation may change the scenario significantly.  NSs with a high degree of differential rotation may be dynamically unstable for  $\beta \gtrsim 0.01$ \cite{Shibata2002,Shibata2003}.  Additionally, an $m=1$ one-armed spiral instability may become unstable if differential rotation is sufficiently strong \cite{Centrella2001,SBM2003}.  

Nevertheless, in practice only very young proto-NSs are affected by differential rotation. In fact, it is expected that a few minutes after a NS is born in a core collapse supernova, rigid rotation sets in due to the presence of viscosity or a sufficiently strong magnetic field \cite{Haensel2016}. Thus, these dynamical instabilities are not thought to have an impact in most cold catalyzed NSs.

Another class of nonaxisymmetric instabilities are  secular instabilities,  which  require the presence of dissipation  due, for example, to viscosity and/or GWs. 
In this context the so-called $r-$modes have attracted considerable attention. They are associated with large-scale currents in NSs that couple to gravitational radiation and remove energy and angular momentum from the star in the form of GWs \cite{Andersson98,FriedmanMorsink,Friedman78,LOM98}. In the absence of viscous dissipation, they are unstable at all rotation frequencies  leading to an exponential rise of the $r-$mode amplitude. 
However, when viscous damping is taken into account the star is stable at low frequencies and instability windows are expected  only at high enough frequencies \cite{LOM98,Alford2012b}. 
In addition, the fact that fast-spinning compact stars are observed suggests that nonlinear damping  mechanisms are present and limit the exponential grow of $r-$modes that would destroy the NS \cite{Alford2012a}. 
If this instability is stopped at a large amplitude, $r-$modes may be a strong and continuous source of GWs in some objects which can be used to probe the NS interior structure.

\section{Summary and Future Prospects}
\label{SVI}

In this work we have presented a general overview of some
  of the hot topics and open issues concerning the behavior of the
  matter in the dense cores of NSs. Particular emphasis has been put on
  the EoS associated with such matter and its connection to NS
  observables constrained by the gravitational wave event GW170817.
  Models for the EoS have been presented which are based on the
  relativistic (density dependent) mean-field approximation to model
  hadronic matter and a nonlocal extension of the three-flavor
  Nambu-Jona-Lasinio model for quark matter.  Assuming that the latter
  is in the 2SC+s phase, we have presented hybrid EoSs with sharp
  quark-hadron interface.  Stable hybrid stars are found to be
  composed of 60-75$\%$ of diquark matter. Motivated by the
  gravitational-wave event GW170817, we show the results obtained for $g$-mode
  frequencies of oscillating NSs \cite{Ranea-Sandoval:2018bgu}. Since these modes depend strongly on
  the amplitude of the energy density discontinuity at the
  quark-hadron interface, $g$-modes may be used to learn about the
  presence and size of the energy density jump that characterized the
  quark-hadron interface.  An analysis of the $f$ and $g$ modes
  suggest that they can be described by fits which are rather
  independent of the nuclear EoS. Aside from constraints derived from
  $f$ and $g$ modes, GW from NSs mergers impose restrictions on the
  masses and radii of NSs, which, in turn, sheds light on their
  interior compositions. In particular, such detections will also be
  very helpful to determine the sought-after upper mass limit of NSs.
  
  Although LIGO's second observing Run O2 ended on august of 2017, it is estimated that the third observing Run O3 will start at the beginning of 2019\footnote{LIGO/Virgo O3 run started 2019 April 1 at 15:00 UTC. In this new era, data is made publicly available to allow for rapid follow-up of events by the electromagnetic astrophysical community (https://gracedb.ligo.org/).}. These new searches for gravitational-waves from stellar-mass coalescing compact binaries will help astrophysicists to provide new constraints on the EoS of dense nuclear matter composing NSs.

In addition, observational efforts are emphasized through several projects to allow for precise determinations of NS masses and radii. Large effective areas combined with high spectral and timing resolutions of the next generation of X-ray instruments such as the already working Neutron Star Interior Composition ExploreR (NICER\footnote{https://www.nasa.gov/nicer}) and future planned ATHENA\footnote{http://sci.esa.int/athena/} will substantially help to improve our current picture. NICER was designed to reach about 5\% accuracy on the determination of the mass and the radius of the closely PSR J0437--4715 by directly fitting the phase-resolved shape of its X-ray pulses. In a recent paper, the authors use NICER observations to update the X-ray timing of radio- and $\gamma$-quiet pulsars PSR J1412+7922 and PSR J1849--0001 \cite{bogdanov2019}. Moreover, five X-ray bursts on 4U 1820--30 were already detected with NICER. An analysis of the first of them has been recently published \cite{Keek:2018jrj}. The high sensitivity of ATHENA will improve radius estimates of isolated NSs, and cooling tails of X-ray bursts, and allow for searches of spectral features in NS atmospheres. At radio wavelengths, the Square Kilometer Array (SKA\footnote{https://www.skatelescope.org/}) will become the largest radio facility, beginning science operations very soon, allowing for the discovery of a significant number of new pulsars, detect hundreds of glitches, and to perform new NS mass determinations.

\section*{Acnowledgements}
This research was supported by Universidad Nacional de La Plata and CONICET
(Argentina), Grants G140, G157, X824 and PIP-0714, and the National Science
Foundation (USA), grant number PHY-1714068. Germán Lugones acknowledges CNPq Brazil for partial financial support.

%\section*{References}

% Override the revtex href command in order that the JHEP bib style
% will work properly:
%\renewcommand{\href}[2]{#2}

% macros used by ADS Database BiBTeX entries:
% see http://adsabs.harvard.edu/abs_doc/aas_macros.sty
\newcommand{\apj}{Astrophys. J.\ }
\newcommand{\prd}{Phys. Rev. D \ }
\newcommand{\prc}{Phys. Rev. C \ }
\newcommand{\apjl}{Astrophys. J. Lett.\ }
\newcommand{\mnras}{Mon. Not. R. Astron. Soc.\ }
\newcommand{\aap}{Astron. Astrophys.\ }
\newcommand{\jcap}{Journal of Cosmology and Astroparticle Physics\ }

\bibliographystyle{JHEP_MGA}
\bibliography{Orsaria_JPG_GW_review}

\end{document}